\documentclass[conference]{IEEEtran}

\usepackage{graphicx}
\usepackage{booktabs}
\usepackage{amsmath}
\usepackage{amsthm}
\usepackage{amssymb}
\usepackage{amsfonts}
\usepackage{epsfig}
\usepackage{graphicx}
\usepackage{subfigure}
\usepackage[bottom]{footmisc}
\usepackage{balance}
\usepackage{algpseudocode}
\usepackage{algorithmicx}
\usepackage{algorithm}
\usepackage{multirow}
\usepackage{xspace}
\usepackage[singlelinecheck=off]{caption}
\DeclareCaptionType{copyrightbox}
\usepackage{pifont}
\newtheorem{theor}{Theorem}

\newtheorem{exam}{Example}
\newtheorem{problem}{Problem}

\newcommand{\spara}[1]{\smallskip\noindent{\bf #1}}

\newcommand{\squishlist}{
 \begin{list}{$\bullet$}
  {  \setlength{\itemsep}{0pt}
     \setlength{\parsep}{3pt}
     \setlength{\topsep}{3pt}
     \setlength{\partopsep}{0pt}
     \setlength{\leftmargin}{2em}
     \setlength{\labelwidth}{1.5em}
     \setlength{\labelsep}{0.5em}
} }
\newcommand{\squishlisttight}{
 \begin{list}{$\bullet$}
  { \setlength{\itemsep}{0pt}
    \setlength{\parsep}{0pt}
    \setlength{\topsep}{0pt}
    \setlength{\partopsep}{0pt}
    \setlength{\leftmargin}{2em}
    \setlength{\labelwidth}{1.5em}
    \setlength{\labelsep}{0.5em}
} }

\newcommand{\squishdesc}{
 \begin{list}{}
  {  \setlength{\itemsep}{0pt}
     \setlength{\parsep}{3pt}
     \setlength{\topsep}{3pt}
     \setlength{\partopsep}{0pt}
     \setlength{\leftmargin}{1em}
     \setlength{\labelwidth}{1.5em}
     \setlength{\labelsep}{0.5em}
} }

\newcommand{\squishend}{
  \end{list}
}

\newcommand{\eat}[1]{}

\newcounter{ccc}

\newcommand{\bigO}{\mathcal{O}}

\begin{document}

\title{Composite Hashing for Data Stream Sketches}

\author{\IEEEauthorblockN{Arijit Khan}
\IEEEauthorblockA{Nanyang Technological University, Singapore\\
arijit.khan@ntu.edu.sg
}
\and
\IEEEauthorblockN{Sixing Yan}
\IEEEauthorblockA{Nanyang Technological University, Singapore\\
alfonso.yan@ntu.edu.sg
}
}

\maketitle

\begin{abstract}
In rapid and massive data streams, it is often not possible to estimate the
frequency of items with complete accuracy. To perform the operation in a reasonable amount
of space and with sufficiently low latency, approximated methods are used. The most common
ones are variations of the {\sf Count-Min} sketch. By using multiple hash functions,
they summarize massive streams in sub-linear space. In reality, data item ids or keys can be
modular, e.g., a graph edge is represented by source and target node ids,
a 32-bit IP address is composed of four 8-bit words, a web address consists of
domain name, domain extension, path, and filename, among many others.
In this paper, we investigate the modularity property of item keys, and
develop more accurate, composite hashing strategies, such as employing multiple independent
hash functions that hash different modules in a key and their combinations separately,
instead of hashing the entire key directly into the sketch.

Our problem of finding the best hashing strategy is non-trivial, since there
are exponential number of ways to combine the modules of a key before they can be hashed into
the sketch. Moreover, given a fixed size allocated for the entire sketch, it is hard
to find the optimal range of all hash functions that correspond to different
modules and their combinations. We solve both problems with 
theoretical analysis, and perform thorough experiments with real-world datasets to demonstrate the
accuracy and efficiency of our proposed method, {\sf MOD-Sketch}.
\end{abstract}

\vspace{-3mm}
\section{Introduction}
\label{introduction}
\vspace{-2mm}
In many domains such as real-time IP traffic, telephone calls, text data from email/SMS/blog/social media,
web clicks and crawls, measurements from sensors and scientific experiments,
rapid and massive volume of data are generated as a stream~\cite{CJKMSS04,MP09}.
In order to process fast streaming data, a growing number of
applications relies on devices such as network interface cards, routers, switches, cell processors,
{\sf FPGA}s, and {\sf GPU}s \cite{RKK14}; and usually these devices have very small on-chip memory.
Whether in specialized hardware or in conventional architectures (e.g., static RAM, such as CPU cache),
efficient processing of fast and large stream data requires creation of a succinct synopses in
a single pass over that stream \cite{CGMP12}. These summaries must be updated incrementally with
incoming items. They should also support online query answering, e.g., estimating the frequency
of an item in a long stream. Due to smaller summary sizes compared to the original stream, it is
often not possible to answer queries with complete accuracy ---
reducing summary size increases the efficiency, but also reduces the accuracy.

\begin{figure}[t!]
\vspace{-2mm}
\centering
\subfigure[{\sf Count-Min}]  {
\includegraphics[scale=0.2]{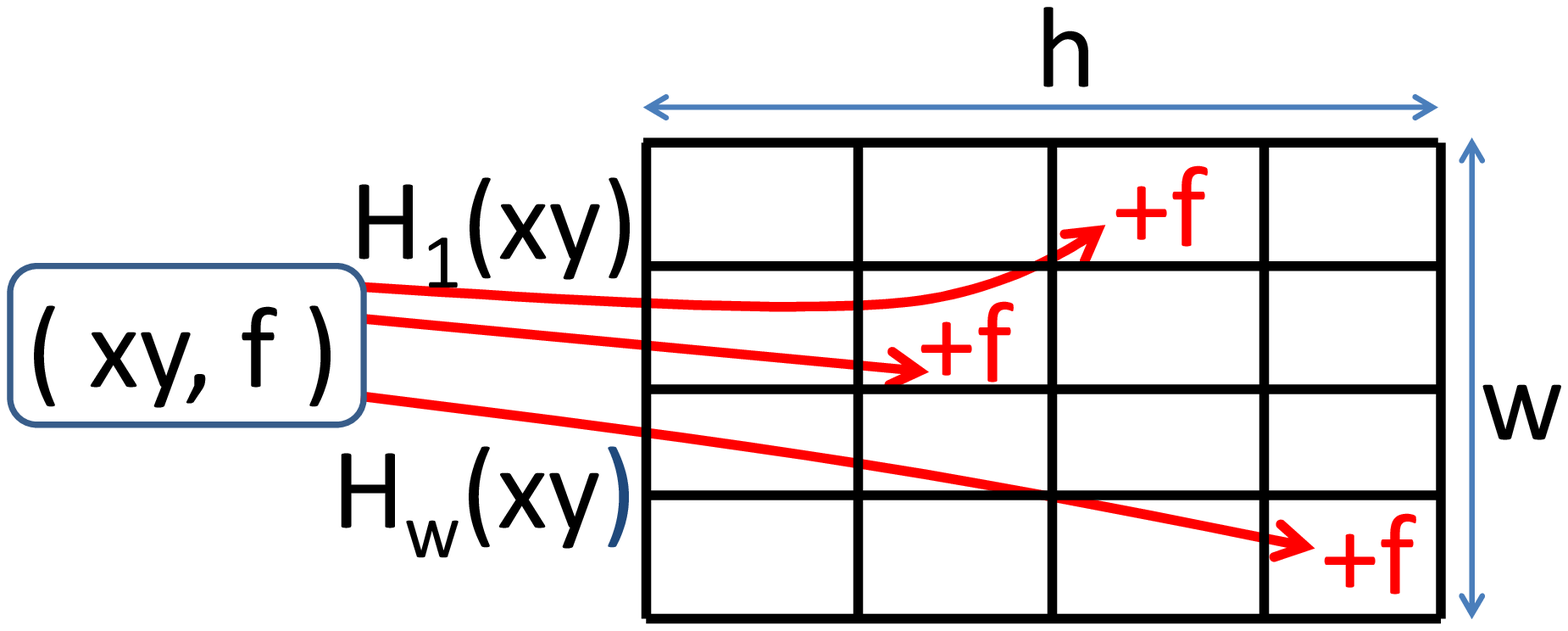}
\label{fig:cmsketch}
}
\subfigure[{\sf MOD-Sketch}]  {
\includegraphics[scale=0.2]{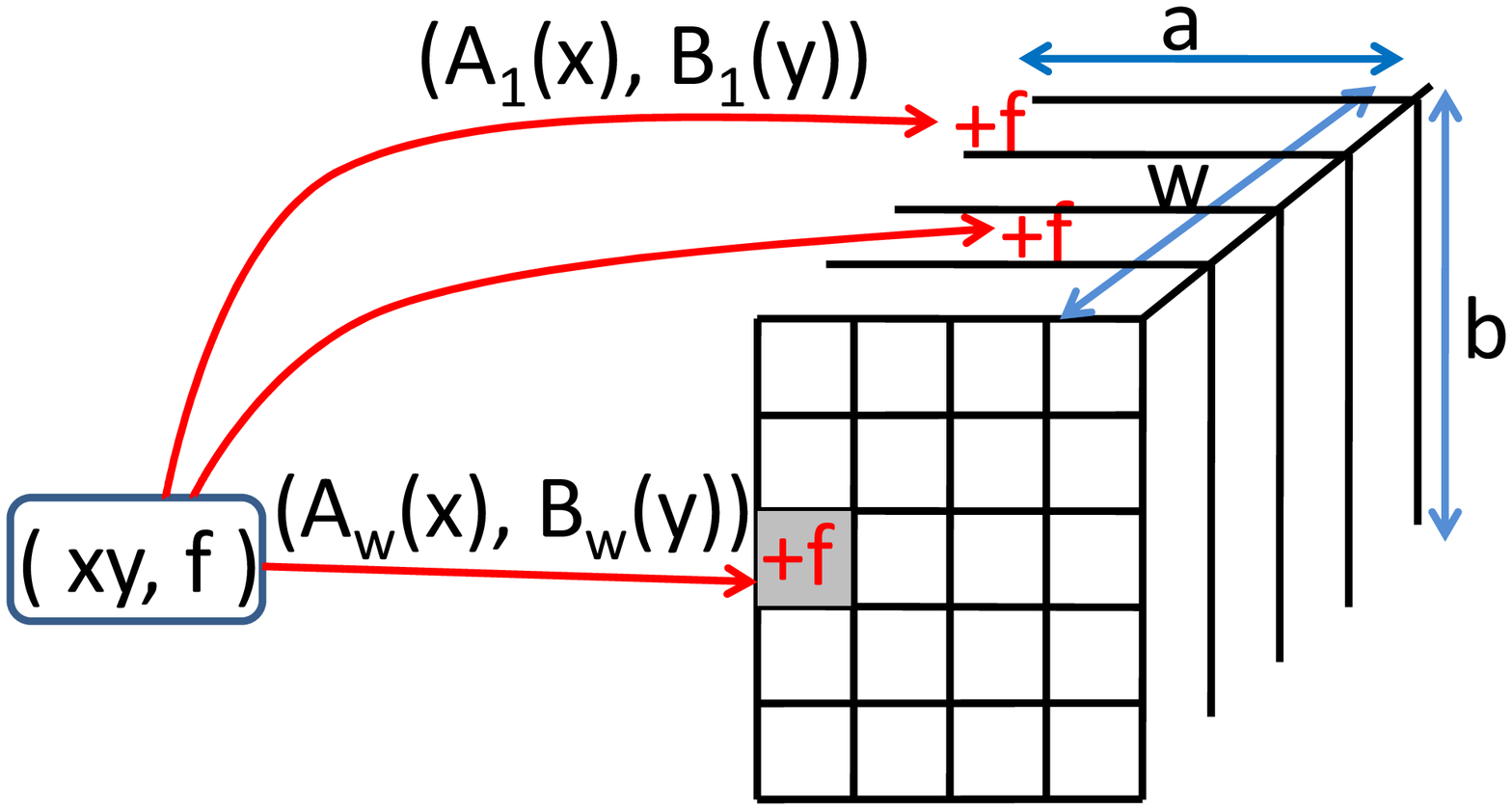}
\label{fig:modsketch}
}
\vspace{-4mm}
\caption{\small {\sf Count-Min} vs. {\sf MOD-Sketch}: {\sf Count-Min} hashes the entire key directly into the sketch, having $w$ hash functions
and each hash function with range $h$. Since the modularity of a key is two in this example, {\sf MOD-Sketch} employs two independent hash functions that hash two modules of the key separately, the first
hash function with range $a$ and the second one with range $b$. In total, {\sf MOD-Sketch} uses $w$ such pairs of hash functions. To utilize
same space, we ensure that $a\times b=h$. In this work, we investigate which design is better, and the optimal values of $a$ and $b$, given $h$.}
\label{fig:cmsketch_and_modsketch}
\vspace{-8mm}
\end{figure}

We study {\em frequency estimation queries} over data streams:
Given a data item as a query, we estimate the frequency count of that
item in the input stream. In the literature, sketch data
structures~\cite{DGGR02,countmin,AMSz96,CCF02,GKMS02,KSZC03}
have been employed for frequency estimation queries. By using multiple hash functions,
they summarize massive data streams within a limited memory space. For example
in Figure~\ref{fig:cmsketch}, we depict the {\sf Count-Min} sketch, having $w$ pairwise
independent hash functions and each hash function with range $h$. The key of every
data item is hashed directly into the sketch.

In various settings, item keys are modular, i.e., consisting of multiple ordered parts.
Many domains of stream data such as graph edges (e.g., web graphs, transportation networks, and social networks),
IP addresses, telephone numbers, web addresses belong to this category. This enables us with more optimization scopes for
hashing, such as one can hash different modules in a key and their combinations separately (i.e., {\em composite hashing}).
In Figure~\ref{fig:modsketch}, we illustrate an item key having modularity two,
and the two parts are hashed via two independent hash functions separately, the first
hash function with range $a$ and the second one with range $b$. To allocate the same amount of space as {\sf Count-Min},
we ensure that $a\times b=h$, and the sketch (referred to as the {\sf MOD-Sketch}, an abbreviation for \underline{mod}ular sketch)
uses $w$ such pairs of hash functions. Two immediate questions that arise are as follows: {\bf (1)} How does one select the optimal
values of $a$ and $b$? {\bf (2)} Between {\sf Count-Min} and {\sf MOD-Sketch}, which one results in more
accuracy?

Our problems are non-trivial --- answers to both questions depend on underlying properties of the stream
and parameters of the sketch. For example, consider the case: $a$=$b$=$\sqrt{h}$, and a stream of graph
edges in the form $(\langle x, y \rangle, f)$. Here, $\langle x, y \rangle$ represents a directed edge
from the source node $x$ to the target node $y$, with frequency $f$. Next, consider the edges having
the same source node. {\sf MOD-Sketch} maps the source node to same hash values, resulting in
more collisions across edges having the same source node. This problem is exacerbated when the number of
source nodes is more than that of target nodes in the graph stream. However,
such skewness also allows us to find better {\sf MOD-Sketch} parameters, e.g., $a>b$.
Similar optimization, on the contrary, is not possible for {\sf Count-Min}.
We find optimal values of $a$ and $b$ in a {\em data-dependent} manner: By sampling \footnotemark[1] a small portion of the stream, and analyzing
out-degrees of source nodes and in-degrees of target nodes in this sample. Furthermore,
we decide the best choice between {\sf Count-Min} and {\sf MOD-Sketch}, again by sampling
a small portion of the stream, and then based on the standard deviation of the values stored
in different cells of these two sketches. We discuss our solutions with theoretical
analyses, and demonstrate their performance with empirical results.

\footnotetext[1]{\scriptsize Sampling a portion of the stream for estimating better sketch parameters
has been considered earlier, e.g., in gSketch \cite{gsketch} and FCM \cite{TBAY09}. It is worth noting
that similar to those works, we assume that a sample of the original stream is available, which
retains its important characteristics, e.g., distribution of high-frequency items.}

After studying the optimal hashing strategy for keys with modularity two, we
focus on generalizations where keys can have higher modularity, e.g., a 32-bit IP
address has modularity four (composed of four 8-bit words), thus opening the stage to
a wider scope of optimizations. This problem is more complex due to two reasons.
{\bf (1)} There are exponential number of ways to combine the modules of a key before
they can be hashed into the sketch. {\bf (2)} Given a fixed $h$, it is hard to find the
optimal ranges of all hash functions that correspond to different modules and their
combinations. 
We propose greedy heuristics, and empirically
demonstrate that our method achieves higher accuracy compared to several baselines.

\vspace{-0.7mm}
\spara{Our contribution and roadmap.} We summarize our contributions below.
\vspace{-1mm}
\begin{itemize}
\item We investigate the fundamental problem of
constructing more accurate, composite hashing for sketches over data streams (\S~\ref{sec:problem_formulation}).
\item We devise scalable, effective, and data-dependent solution for finding the optimal hashing
ranges for different modules of an item key, with theoretical analyses (\S~\ref{sec:opt_two}).
\item We further generalize our algorithm to find good-quality hashing strategies
for keys with modularity higher than two (\S~\ref{sec:alg}).
\item We perform detailed experiments to demonstrate the accuracy, efficiency, and throughput
of our developed {\sf MOD-sketch}, comparing it with several baselines including {\sf Count-Min}
\cite{countmin} and {\sf Equal-Sketch} \cite{KA17,TCM16}, while also demonstrating its
generalizability by implementing {\sf MOD-Sketch} on top of the {\sf FCM} sketch \cite{TBAY09} (\S~\ref{sec:experiments}).
\end{itemize}

\vspace{-2mm}
\section{Related Work}
\label{sec:related}
\vspace{-2mm}
\spara{Data stream sketches.}
The problem of synopsis construction has been studied extensively~\cite{CGMP12} in the context of a variety of techniques
such as sampling~\cite{GGMS96,C15} (including graph stream sampling, e.g., \cite{PTTW13,ADWR17}), wavelets~\cite{GSWS01,GG02}, histograms~\cite{GKS01},
sketches~\cite{DGGR02,countmin,AMSz96,CCF02,GKMS02,KSZC03}, and counter-based methods,
e.g., {\sf Space Saving}~\cite{MAA05} and {\sf Frequent}~\cite{CH08}. Sketches and counter-based approaches are
widely used for stream data summarization. Sketches are typically used for {\em frequency estimation}, which is the focus of this paper.
Sketches keep approximate counts for all items, counter-based approaches maintain approximate counts only for the top-$k$ frequent items.

Among various sketches \cite{countmin,AMSz96,CCF02,EV03,FSGMU98,GKMS02,KSZC03},
{\sf Count-Min}~\cite{countmin} is widely studied, it achieves good update throughput in general, as well as very high accuracy on
skewed distributions. Several approaches have been proposed to further improve its accuracy,
e.g., frequency-aware hashing ({\sf FCM} \cite{TBAY09}, {\sf gSketch} \cite{gsketch}, and {\sf ASketch} \cite{RKA16}),
and non-uniform counter sizes ({\sf Cold Filter} \cite{ZYJCYLU18}). However, they do not consider the
modularity of item keys unlike ours.

In the domain of graph edge stream, {\sf Count-Min} has
been extended to {\sf gMatrix} \cite{KA17} and {\sf TCM} \cite{TCM16},
which separately hash the source and target node ids corresponding to
an edge. These approaches follow composite hashing. However, unlike ours,
they allocate the same hashing range to both source and target nodes. Moreover, they do not address
the problem of hashing data items with modularity higher than two.
Composite hashing over IP data (i.e., having modularity four) is discussed in \cite{SLCGGZDKM06}.
Once again, they allocate the same hashing range to all 4 bytes of the IP address.
For generality, in this paper we refer to these prior works as {\sf Equal-Sketch}. Based on detailed experiments,
our proposed {\sf Mod-Sketch} results in higher accuracy compared to {\sf Equal-Sketch}.

\spara{Data-dependent and other composite hashing.} Recently, data-driven learning methods for advanced hash functions (i.e., learning to hash)
have become popular, particularly in the context of {\em nearest neighbor} queries (for a survey, see \cite{WLKC16}). Composite hashing for nearest neighbor
queries has been studied in \cite{ZWS11}. While we select ranges of our composite hash functions in a data-dependent manner, our focus
is {\em sketches} for data stream summarization. This is different from existing work on learning to hash for nearest neighbors. 
\vspace{-1mm}
\section{Preliminaries}
\label{sec:problem_formulation}
\vspace{-2mm}
The incoming data stream contains tuples $(i_1, f_1)$, $(i_2, f_2)$,
$\ldots$, $(i_t,$ $f_t)$, $\ldots$. Here, $(i_t, f_t)$ denotes the arrival of the $t$-th tuple
with item $i_t$ having an associated positive count $f_t$. In many applications, the value of $f_t$ is set to one,
though we assume an arbitrary positive count in order to retain the generality of our model. As an example,
in a telecommunication application, the frequency count $f_t$ may denote the number of seconds in the $t$-th phone conversation.
An item may appear multiple times in the stream, i.e., it is possible that $i_{t}=i_{t'}$,
for $t \ne t'$. When we issue a query, e.g., finding the frequency of an item $i$, we are looking for
the aggregate count of that item in the stream so far. While sketch-based methods including ours can be adapted for
time-window queries \cite{Aggarwal03}, as well as for removal of items \cite{CH08} (i.e., a
negative-count-update can be performed in the same way as a positive-count-update, so long as the
overall count of an item never becomes negative), we do not consider them in this work. Furthermore,
we use the notation $i$ interchangeably to denote an item as well as its key, because a key uniquely
identifies an item.

An item key is often modular, that is, composed of $n$ ordered parts: $\langle x^{(1)}, x^{(2)},$ $\ldots, x^{(n)}\rangle$.
Here, $x^{(j)}$ denotes the generic representation of the $j$-th module (i.e., co-ordinate) of a key.
For an item key $i$, we represent its modules as: $i=\langle x_i^{(1)}, x_i^{(2)}, \ldots, x_i^{(n)}\rangle$
Clearly, $x_i^{(j)}$ denotes the specific value in the $j$-th module for item $i$. As an example, in case of
telecommunication network, each item is a communication (edge) from source to target users (nodes).
The domains of source and target nodes are generally known apriori, and the edges between them arrive
continuously in the stream. {\em In case of modular keys, we assume that each module $x_i^{(j)}$ of a key $i$
can be from a predefined set of integers.} This holds in reality, e.g., for a 32-bit
IP address which consists four 8-bit words, each word can be from 000 to 255.

In this paper, we shall discuss our framework, {\sf MOD-Sketch} (an abbreviation for \underline{mod}ular sketch)
as a composite hashing strategy on top of the {\sf Count-Min}, one of the most widely-studied sketches.
Nevertheless, {\em the {\sf MOD-Sketch} framework is generic, and it can be applied in
combination with several other sketches. In our experiments in \S~\ref{sec:general}, we also demonstrate the performance
of {\sf MOD-Sketch} when it is constructed with other underlying sketches}.
Next, for ease in the presentation, we introduce the standard {\sf Count-Min} sketch.
\vspace{-5mm}
\subsection{Count-Min}
\vspace{-1mm}
In {\sf Count-Min}, a hashing approach is employed to
approximately maintain the frequency counts of a large number of distinct items
in a stream (Figure~\ref{fig:cmsketch}). We use $w= \lceil \mbox{ln}(1/\delta) \rceil$
pairwise independent hash functions, each of which maps onto
uniformly random integers in the range $h=[0, e/\epsilon]$, where
$e$ is the base of the natural logarithm, $\epsilon$ and $\delta$ are terms
to define the error and the probabilistic error guarantee, respectively,
which we shall introduce shortly.
The data structure
consists of a $2$-dimensional array with $h \times w$ cells of length
$h$ and width $w$. Each hash function corresponds to one of $w$
$1$-dimensional arrays with $h$ cells each. Next, consider a data stream
with items drawn from a massive set of domain values. When an item
is received, we apply each of the $w$ hash functions to map onto a number
in $[0 \ldots h-1]$. The count
of each of these $w$ cells is incremented by $1$.  To
{\em estimate} the count of an item, we determine the set of $w$
cells to which each of the $w$ hash-functions maps, and compute the
minimum value among all these cells. Let $c_t$  be the true value of
the count being estimated. We note that the estimated count is at
least equal to $c_t$, since we are dealing with non-negative counts
only, and there may be an over-estimation because of collisions
in hash cells. It has been shown in
\cite{countmin} that for a data stream with $L$ arrivals, the
estimate is at most $c_t + \epsilon \cdot L$ with probability at
least $1 - \delta$. In the event that the items have frequencies
associated with them, we increment the corresponding count with the
appropriate frequency. The same bounds hold in this case, except
that we define $L$ as the sum of the frequencies of the items
received so far.

For {\sf Count-Min} to be effective, the hash functions are required
to be pairwise independent. Following the {\sf Count-Min} work \cite{countmin},
we select modular hash functions as follows.
\vspace{-2mm}
\begin{equation}
H(i)=\left((q\times i + r) \mod P \right) \mod h
\vspace{-2mm}
\label{equ:mod_hash}
\end{equation}
Here, $P$ is a prime number larger than the maximum value of any key id $i$, and $h$ is the range of the hash function.
We select $q$ and $r$ uniformly at random from sets $\{1,2,\ldots,P-1\}$ and  $\{0,1,\ldots,P-1\}$, respectively.
\vspace{-2mm}
\subsection{Hashing Keys with Modularity Two}
\vspace{-1mm}
We next introduce our problem for hashing keys with modularity two, i.e., $i=\langle x_i^{(1)}, x_i^{(2)}\rangle$.
The hashing problem for higher modularity keys will be discussed in \S~\ref{sec:alg}.

As we demonstrated in Figure~\ref{fig:cmsketch_and_modsketch}, there are two possible choices for hashing the keys
with modularity two. {\bf (1)} Concatenate two ordered integer modules, construct a single integer id, i.e., $i=x_i^{(1)}x_i^{(2)}$,
and hash it directly in the {\sf Count-Min}. To distinguish between the keys such as $(1,12)$ and $(11,2)$,
we first consider the domains of the modules before concatenating them. For example, if the domain of each module
is the set of integers $\in (0,99)$, then $(1,12)$ is concatenated as $0112$, whereas $(11,2)$ is concatenated as $1102$.
{\bf (2)} Employ two independent hash functions that hash two modules of the key separately
in {\sf MOD-Sketch}. In total, {\sf MOD-Sketch} uses $w$ such pairs of hash functions. All $2w$
hash functions need to be pairwise independent, which is ensured by modular hash functions as in Equation~\ref{equ:mod_hash}.
Let the range of the hash functions in {\sf Count-Min} be $h$,
whereas for {\sf MOD-Sketch} the ranges are $a$ and $b$. To ensure the same amount of space, we have: $a\times b = h$.
The other parameter, i.e., the number of hash functions, $w$ remains the same for both sketches, as the probability of the error bound is
determined by $w$, whereas the actual error bound depends on the range of the hash functions: $h$, $a$, $b$.

\spara{Probabilistic accuracy guarantee.} We now derive the accuracy guarantees for {\sf Count-Min} and {\sf MOD-Sketch}.
It is important to note that one may not directly compare the error bounds of these two sketches, since
the hashing strategies are different (i.e., {\sf Count-Min} hashes concatenated keys, whereas {\sf MOD-Sketch} performs
composite hashing). In \S~\ref{sec:opt_mod_cm}, we shall introduce a more practical approach to select the best choice between them for a specific
data stream.
\begin{theor}
Let the total frequency of items received so far in the stream be denoted by $L$. Let $\overline{Q(x_i^{(1)},x_i^{(2)})}$ be the true frequency
of the item $i=(x_i^{(1)},x_i^{(2)})$. Let $\epsilon \in (0,1)$ be a very small fraction.
Consider {\sf Count-Min} with hash function range $h$ and width $w$. Then, with probability at least $1-(1/h\epsilon)^{w}$,
the estimated frequency $Q(x_i^{(1)},x_i^{(2)})$ is related to the true frequency by the following relationship:
\vspace{-2mm}
\begin{align}
& \overline{Q(x_i^{(1)},x_i^{(2)})} \leq Q(x_i^{(1)},x_i^{(2)}) \leq \overline{Q(x_i^{(1)},x_i^{(2)})}+L\epsilon &
\label{equ:cmhash_error}
\vspace{-4mm}
\end{align}
\label{th:cmhash_error}
\end{theor}
\vspace{-4mm}
The proof of Theorem~\ref{th:cmhash_error} can be found in the Count-Min paper \cite{countmin}. In Equation \ref{equ:cmhash_error},
the error term is relatively small if the true frequency $\overline{Q(x_i^{(1)},x_i^{(2)})}$ is a significant fraction of the aggregate
frequency $L$. For real-world streams, which often has a skew \cite{MP09}, this holds for the high-frequency items.
\vspace{-1mm}
\begin{theor}
Let the total frequency of items received so far in the stream be denoted by $L$. Let $\overline{Q(x_i^{(1)},x_i^{(2)})}$ be the true frequency
of the item $i=(x_i^{(1)},x_i^{(2)})$. Let $O(x_i^{(1)},*)$ be the sum of frequencies of the items having the first module as $x_i^{(1)}$, and $O(*,x_i^{(2)})$
the sum of frequencies of the items having the second module as $x_i^{(2)}$. Let $\epsilon \in (0,1)$ be a very small fraction.
Consider {\sf MOD-Sketch} with hash function ranges $a$ and $b$, as well as width $w$. Then, with probability at least $1-(3/ab\epsilon)^{w}$,
the estimated frequency is related to the true frequency by the following relationship:
\vspace{-1mm}
\begin{align}
& \overline{Q(x_i^{(1)},x_i^{(2)})} \leq Q(x_i^{(1)},x_i^{(2)}) \leq \overline{Q(x_i^{(1)},x_i^{(2)})}& \nonumber \\
& \qquad +\left[L + O(*,x_i^{(2)})\cdot b + O(x_i^{(1)},*)\cdot a\right]\cdot \epsilon  &
\label{equ:mod_hash_error}
\vspace{-2mm}
\end{align}
\label{th:mod_hash_error}
\end{theor}
\vspace{-7mm}
\begin{proof}
Any incoming item,
for which the ordered modules are neither $x_i^{(1)}$ nor $x_i^{(2)}$, is equally
likely to map onto one of $ab$ cells of a hash function. The probability that any incoming item
maps onto a particular cell is given by $1/ab$. Therefore, the expected number of spurious items
for which the modules are neither $x_i^{(1)}$ nor $x_i^{(2)}$, yet they get mapped onto the cell
$(A_{k}(x_i^{(1)}),B_{k}(x_i^{(2)}),k)$ is given by $L/(ab)$.
Let the number of such spurious items for the $k$-th hash function be denoted by the random variable $R_{k}$.
Then, by the Markov inequality:
\vspace{-2mm}
\begin{align}
& P(R_{k}\geq L\cdot \epsilon)\leq E[R_{k}]/(L\cdot \epsilon) \leq 1/(ab\epsilon) &
\vspace{-2mm}
\end{align}
Next, we examine the case of spurious items for which the first module
is $x_i^{(1)}$. The number of such items is $O(x_i^{(1)},*)$ and the expected number of such items
which map onto the entry $(A_{k}(x_i^{(1)}), B_{k}$ $(x_i^{(2)}),k)$  is given by $O(x_i^{(1)},*)/b$. Let $U^{(1)}_{k}$
be the random variable representing the number of such items. Then, by using the Markov inequality, we get:
\vspace{-2mm}
\begin{align}
& P(U^{(1)}_{k}\geq O(x_i^{(1)},*)\cdot a\cdot\epsilon) \leq E[U^{(1)}_{k}]/(O(x_i^{(1)},*)\cdot a\cdot \epsilon)& \nonumber \\
& \leq 1/(ab\epsilon) &
\vspace{-7mm}
\end{align}
Similarly, we denote by the random variable $U^{(2)}_{k}$ the number of items for which the second module
is $x_i^{(2)}$. The number of such items is $O(*,x_i^{(2)})$ and the expected number of such items
which map onto the entry $(A_{k}(x_i^{(1)}), B_{k}$ $(x_i^{(2)}),k)$  is given by $O(*,x_i^{(2)})/a$.
With Markov inequality, we have:
\vspace{-2mm}
\begin{align}
& P(U^{(2)}_{k}\geq O(*,x_i^{(2)})\cdot b\cdot \epsilon) \leq E[U^{(2)}_{k}]/(O(*,x_i^{(2)})\cdot b\cdot \epsilon)& \nonumber \\
& \leq 1/(ab\epsilon) &
\vspace{-7mm}
\end{align}
By combining the three above inequalities with the following rule $P(A \cup B \cup C) \le P(A) + P(B) + P(C)$, we get:
\vspace{-2mm}
\begin{align}
& P\left(R_{k}+U^{(1)}_{k}+U^{(2)}_{k} \geq L\cdot\epsilon+O(*,x_i^{(2)})\cdot b\cdot\epsilon \right.& \nonumber \\
& \qquad \quad \left.+O(x_i^{(1)},*)\cdot a\cdot \epsilon\right) \leq 3/(ab\epsilon) &
\label{equ:mod_hash_bound}
\vspace{-7mm}
\end{align}
Taking the smallest estimate gives the best estimator, and the probability that this estimate
violates the inequality in Equation~\ref{equ:mod_hash_bound} is the probability that all $w$ estimates exceed this error.
The probability that this is true is given by at most $1-(3/ab\epsilon)^w $. The result follows.
\end{proof}
\vspace{-2mm}
For the above probability to be less than 1, we require $ab > 3/\epsilon$. As earlier, since $w$ occurs in the exponent,
the robustness of the above result can be magnified for modest values of $w$.

\vspace{-2mm}
\subsection{Problem Statement}
\vspace{-1mm}
We state our problems for hashing keys with modularity two as follows.
\vspace{-1mm}
\begin{problem}
For a data stream and a pre-defined length $h$, find the most accurate {\sf MOD-Sketch} range parameters $a$ and $b$ such that $a\times b=h$.
\label{prob:ab}
\end{problem}
\vspace{-4mm}
\begin{problem}
For a data stream and a pre-defined length $h$, select the most accurate sketch between {\sf Count-Min} and {\sf MOD-Sketch} having the same size, i.e., $a\times b = h$.
\label{prob:best_sketch}
\end{problem}
\vspace{-1mm}
These are difficult problems as the entire stream may not be available for such computations, and there are several choices for $a$ and $b$ (satisfying $a \times b =h$). We discuss our algorithms for solving Problems~\ref{prob:ab} and \ref{prob:best_sketch} in the next section.

\vspace{-2mm}
\section{Algorithms for Modularity Two}
\label{sec:opt_two}
\vspace{-1mm}
\subsection{Finding High-Quality MOD-Sketch Parameters}
\label{sec:opt_ab}
\vspace{-1mm}
We set to find high-quality hashing ranges $a$ and $b$
for {\sf MOD-Sketch}, by comparing its accuracy guarantee with that of a baseline technique.
For the baseline method, we consider a special version of {\sf MOD-Sketch}, referred to as the
{\sf Equal-Sketch}, where $a=b=\sqrt{h}$. {\em Given a fixed $h$ and for a data stream, we select
$a$ and $b$ such that the error of {\sf MOD-Sketch} is as small as possible, compared
to the error produced by {\sf Equal-Sketch}}. We note that the approach itself does not ensure
finding optimal values of $a$ and $b$. However, based on our empirical results,
hashing ranges $a$ and $b$ found in this manner are of high-quality, and the accuracy of
{\sf MOD-Sketch}, in fact, becomes comparable to that of an exhaustive method which experimentally finds
the best choice of hash function ranges corresponding to two modules of the key.
%
\begin{theor}
Consider an item $i=(x_i^{(1)},x_i^{(2)})$ with its estimated frequency via {\sf MOD-Sketch} and {\sf Equal-Sketch} as
$Q(x_i^{(1)},x_i^{(2)})$ and $Q_E(x_i^{(1)},x_i^{(2)})$, respectively. Let $O(x_i^{(1)},*)$ be the sum of frequencies of the items
having the first module as $x_i^{(1)}$, and $O(*,x_i^{(2)})$ the sum of frequencies of the items having the second module as $x_i^{(2)}$.
We denote by $\alpha$ be the ratio: $\alpha=O(x_i^{(1)},*)/O(*,x_i^{(2)})$, and by $\beta$ the ratio: $\beta=a/b$.
Let $\epsilon \in (0,1)$ be a very small fraction. Then, with a high probability,
$Q(x_i^{(1)},x_i^{(2)})$ is smaller than $Q_E(x_i^{(1)},x_i^{(2)})$ by the largest margin when $\beta=1/\alpha$.
Formally,
\vspace{-2mm}
\begin{align}
& P\left(Q_E(x_i^{(1)},x_i^{(2)})-Q(x_i^{(1)},x_i^{(2)}) \leq O(*,x_i^{(2)})(\sqrt{h}-b)\epsilon\right. \nonumber & \\
& \qquad \quad \left.+ O(x_i^{(1)},*)(\sqrt{h}-a)\epsilon\right) \geq 1-2(3/ab\epsilon)^w &
\label{equ:opt_ab}
\vspace{-2mm}
\end{align}
The difference of $Q(x_i^{(1)},x_i^{(2)})$ from $Q_E(x_i^{(1)},x_i^{(2)})$ is maximized when  $\beta=1/\alpha$.
\label{th:opt_ab}
\end{theor}
\vspace{-3mm}
\begin{proof}
We denote by $\overline{Q(x_i^{(1)},x_i^{(2)})}$ the true frequency of the item $i=(x_i^{(1)},x_i^{(2)})$.
Let the total frequency of items received so far in the stream be $L$.
From Theorem~\ref{th:mod_hash_error}, we get:
\vspace{-2mm}
\begin{align}
& P\left(Q_E(x_i^{(1)},x_i^{(2)})-\overline{Q(x_i^{(1)},x_i^{(2)})} \geq L\epsilon \right. \nonumber & \\
&\left.\qquad + O(*,x_i^{(2)})\sqrt{h}\epsilon + O(x_i^{(1)},*)\sqrt{h}\epsilon\right) \leq (3/ab\epsilon)^w &
\label{in:eq}
\vspace{-7mm}
\end{align}
\vspace{-2mm}
\begin{align}
& P\left(Q(x_i^{(1)},x_i^{(2)})-\overline{Q(x_i^{(1)},x_i^{(2)})} \geq L\epsilon + O(*,x_i^{(2)})b\epsilon \right.\nonumber & \\
&\left.\qquad + O(x_i^{(1)},*)a\epsilon\right) \leq (3/ab\epsilon)^w &
\label{in:mod}
\vspace{-9mm}
\end{align}

\vspace{-2mm}
The inequality in Equation~\ref{in:mod} can be rewritten as follows.
\vspace{-2mm}
\begin{align}
& P\left(-Q(x_i^{(1)},x_i^{(2)})+\overline{Q(x_i^{(1)},x_i^{(2)})} \geq -L\epsilon - O(*,x_i^{(2)})b\epsilon \right.\nonumber & \\
&\left.\qquad - O(x_i^{(1)},*)a\epsilon\right) \leq 1-(3/ab\epsilon)^w &
\label{in:mod_rewrite}
\vspace{-9mm}
\end{align}
By combining two inequalities in Equations~\ref{in:eq} and \ref{in:mod_rewrite} via the rule $P(A \cup B) \le P(A) + P(B)$, we derive:
\vspace{-2mm}
\begin{align}
& P\left(Q_E(x_i^{(1)},x_i^{(2)})-Q(x_i^{(1)},x_i^{(2)})\geq O(*,x_i^{(2)})(\sqrt{h}-b)\epsilon \right. \nonumber & \\
&\left.\qquad + O(x_i^{(1)},*)(\sqrt{h}-a)\epsilon\right) \leq 1-2(3/ab\epsilon)^w &
\vspace{-4mm}
\end{align}
Furthermore, it can be shown with simple calculation that the difference $O(*,x_i^{(2)})(\sqrt{h}-b)\epsilon + O(x_i^{(1)},*)(\sqrt{h}-a)\epsilon$
is maximized when $\beta=a/b=1/\alpha$, where $\alpha=O(x_i^{(1)},*)/O(*,x_i^{(2)})$. Hence, the theorem.
\end{proof}
\vspace{-2mm}
Given a pre-defined $h$ and a data item in the stream, Theorem~\ref{th:opt_ab} helps us to compute high-quality values of $a$ and $b$ for {\sf MOD-Sketch}.
For example, if we observe that $O(*,x_i^{(2)})$ is two times larger than $O(x_i^{(1)},*)$, then we estimate $\beta$ as $2/1$. 
Hence, compared to an {\sf Equal-Sketch} with $a=b=600$, the {\sf MOD-Sketch}
with $a=848$ and $b=424$ is expected to produce the least amount of error. {\em Our result is intuitive, since $O(x_i^{(1)},*)<O(*,x_i^{(2)})$
suggests that there are generally large number of distinct source nodes than that of distinct target nodes, and this results in $a>b$}.

In Theorem~\ref{th:opt_ab}, we compute a high-quality value of $\beta=a/b$ for a specific item $i=(x_i^{(1)},x_i^{(2)})$ in the stream.
For a data stream, we assume that an apriori stream sample is available \cite{gsketch,TBAY09}, which
retains important characteristics of the original stream, e.g., distribution of high-frequency items.
For every item present in the sample, we estimate its related $\alpha=O(x_i^{(1)},*)/O(*,x_i^{(2)})$ from the sample. Next, we employ an aggregate operator (e.g.,
minimum, maximum, median, average, etc.) to compute the aggregated $\alpha_{agg}$. We finally derive $\beta$ as $1/\alpha_{agg}$.
We illustrate our method with an example below. Based on our detailed empirical evaluation over several real-world stream datasets, 
we find that about 2$\sim$4\% initial stream sample, together with the median aggregate of $\alpha$, is generally sufficient to estimate a high-quality value of $\beta$.
\vspace{-1mm}
\begin{exam}
Assume from the initial stream sample, we obtained $3$ items: $(1,2), (1,3), (2,3)$ with aggregated frequency $13, 5, 7$, respectively. We compute
$\alpha$ values for these items: $\alpha(1,2)=O(1,*)/O(*,2)=18/13$. Analogously, we find $\alpha(1,3)=18/12$, $\alpha(2,3)=7/12$.
We compute the median of these $\alpha$ values, i.e., median of the multi-set consisting of $18/12$ for $5$ times, $18/13$ for $13$ times,
and $7/12$ for $7$ times. Thus, $\beta=1/\alpha_{agg}=13/18$.
\end{exam}

\vspace{-4mm}
\subsection{Choice between MOD-Sketch \& Count-Min}
\label{sec:opt_mod_cm}
\vspace{-1mm}
We next consider Problem~\ref{prob:best_sketch}: For a data stream and a pre-defined length $h$,
select the most accurate sketch between {\sf Count-Min} and the optimal {\sf MOD-Sketch} having the same size, i.e., $a\times b = h$.
Unlike Theorem~\ref{th:opt_ab}, we may not directly compare the error bounds of these two sketches, since
the hashing strategies are different. We design a more practical approach by computing the standard deviation of the count values stored in
different cells of these sketches. Theorem~\ref{th:opt_mod_cm} states that the sketch with the smaller standard
deviation generally results in less error for the frequency estimation.
\vspace{-3mm}
\begin{theor}
Consider two sketches $S_1$ and $S_2$ having the same size: $h\times w$, such that the standard
deviations of count values stored in different cells of these sketches
are $\sigma_1$ and $\sigma_2$, respectively. Assume $\sigma_2>\sigma_1$. Let us consider an item $i$
with its estimated frequency via $S_1$ and $S_2$ as $Q_1(i)$ and $Q_2(i)$, respectively.
Assume $\delta >0$. Then, with probability at least $1-2/(1+\delta^2)$, the following holds:
\vspace{-1mm}
\begin{equation}
Q_1(i) - Q_2(i) \leq (\sigma_{1}-\sigma_{2})\cdot \delta
\label{equ:opt_mod_cm}
\vspace{-2mm}
\end{equation}
\label{th:opt_mod_cm}
\end{theor}
\vspace{-5mm}
\begin{proof}
Let the total frequency of items received so far in the stream be $L$.
Let the values stored in the cells of $S_{1}$ be denoted by the random variable $R_{1}$.
The expected value of $R_{1}$ is $L/h$, which is the mean value of the counts in the sketch.
By using the Cantelli's inequality, with $\delta>0$, the possibility of the lower bound of $R_{1}$ can be measured
as follows.
\vspace{-2mm}
\begin{equation}
P(R_{1}-L/h\geq \sigma_{1} \cdot \delta) \leq 1/(1+\delta^{2})
\label{eq:cant1}
\vspace{-2mm}
\end{equation}
Analogously, let the values stored in the cells of $S_{2}$ be denoted by the random variable $R_{2}$.
The expected value of $R_{2}$ is $L/h$.
By employing the Cantelli's inequality, we have:
\vspace{-2mm}
\begin{equation}
P(R_{2}-L/h\geq \sigma_{2} \cdot \delta) \leq 1/(1+\delta^{2})
\vspace{-2mm}
\end{equation}
The above inequality can be written as follows.
\vspace{-2mm}
\begin{align}
& P(R_{2}-L/h\leq \sigma_{2} \cdot \delta) \geq 1- 1/(1+\delta^{2}) \nonumber & \\
\implies &P(-R_{2}+L/h\geq -\sigma_{2} \cdot \delta) \geq 1- 1/(1+\delta^{2}) &
\label{eq:cant2}
\vspace{-2mm}
\end{align}
Combining two inequalities in Equations \ref{eq:cant1} and \ref{eq:cant2}, we get:
\vspace{-2mm}
\begin{equation}
P(R_{1}-R_{2}\leq (\sigma_{1}-\sigma_{2})\cdot \delta) \geq 1- 2/(1+\delta^{2})
\label{equ:opt_mod_cm_prob}
\vspace{-2mm}
\end{equation}
This completes the proof.
\end{proof}
\vspace{-2mm}
Hence, by comparing the variance of two sketches with the same $h$ and $w$ (but
with different hashing techniques), one can predict which sketch would result in less frequency estimation
error for the given data stream.
In practice, we consider an initial sample of the incoming stream, and store it in the two sketches having same $h\times w$. 
Then, based on the standard deviation of the values stored in different cells of these sketches, we decide which sketch we shall use.

\vspace{-2mm}
\section{Algorithms for Modularity $>$ Two}
\label{sec:alg}
\vspace{-1mm}
In this section, we consider the generalization of our problem
for item keys with modularity higher than two.
Two natural baselines for hashing such keys are:
{\bf (1)} {\sf Count-Min}, i.e., concatenate $n$ ordered
modules of the key and hash it directly with a hash
function of range $h$, and {\bf (2)} {\sf Equal-Sketch}, i.e.,
separately hash $n$ modules with $n$ independent hash functions, each
having equal range $(h)^{1/n}$. However, a more accurate {\sf MOD-Sketch}
could combine some modules of the key and hash them together, and
could separately hash the remaining modules. Therefore,
it raises the problem of how to design a more accurate {\sf MOD-Sketch}
for a given key with modularity higher than two, considering modules
might be combined or kept separate as necessary.

\vspace{-3mm}
\subsection{An Exact Strategy}
\label{sec:exact}
\vspace{-1mm}
Let us consider keys of modularity $n$, i.e., $\{x^{(1)},x^{(2)},\ldots,x^{(n)}\}$.
There are exponentially large number of ways
to combine the modules --- also known as the ``Bell number'',
which counts the possible partitions of a set.
We denote by $T(n)$ the total number of ways one can combine different parts of a key
having modularity $n$. Then, we have:
\vspace{-1mm}
\begin{equation}
\displaystyle T(n)=\sum_{k=0}^{n-1}\left[^{(n-1)}{\mathbb C}_k \cdot T(n-k-1)\right]
\end{equation}
with base cases: $T(0)=T(1)=1$. $^{(n-1)}{\mathbb C}_k=\frac{(n-1)!}{k!(n-k-1)!}$.
\begin{table} [tb!]
\centering
\scriptsize
\vspace{-2mm}
\caption{\small $T(n)$ denotes \#ways the modules of a key with modularity $n$ can be combined.
$T(n)$ increases at a higher rate than $2^n$.}
\label{tab:ways}
\vspace{-2mm}
\begin{tabular} { l|l|l|l|l|l|l|l|l|l }
$n$              & $2$  & $3$ & $4$ & $5$ & $6$ & $7$ & $8$ & $9$ & $10$  \\ \hline \hline
$T(n)$           & 2    &  5  &  15 & 52  & 203 & 877 & 4140&21147& 115975 \\ \hline
$2^n$            & 4    &  8  &  16 & 32  & 64  & 128 & 256 & 512 & 1024  \\
\end{tabular}
\vspace{-4mm}
\end{table}
\begin{figure}[tb!]
\centering
\includegraphics[scale=0.24]{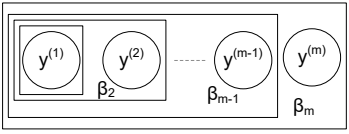}
\vspace{-2mm}
\caption{\small Computation of range ratio $\beta$ for keys with modularity $>2$.}
\label{fig:ratioModel}
\vspace{-6mm}
\end{figure}
\vspace{-2mm}
\begin{proof}
We shall assume that $n>1$, since the base cases for $n=0$ and $n=1$ are already given.
Let us consider the first module in order, i.e., $x^{(1)}$. Clearly, the number of ways $x^{(1)}$
remains as a separate module is $T(n-1)$. Next, we consider the number of ways $x^{(1)}$ can
be combined with exactly one of the remaining modules. The number of remaining modules is
$(n-1)$, and after $x^{(1)}$ is combined with exactly one of the remaining modules, e.g.,
$x^{(1)}x^{(2)}$, we have total $T(n-2)$ possible ways. In general, when $x^{(1)}$ is
combined with exactly $k$ of the remaining modules, where $1\leq k \leq n-1$, total number of possible
ways is $^{(n-1)}{\mathbb C}_k \cdot T(n-k-1)$. By combining all these counts as they are mutually exclusive,
the result follows.
\end{proof}

Notice that for $n >4$, $T(n)>2^n$ (Table \ref{tab:ways}). We observe that {\em $T(n)$, in fact,
increases at a much higher rate than $2^n$}. Therefore, an exact method to verify all possible ways of combining
the modules and thereby finding the best option would be extremely time consuming.
In this paper, we keep open the complexity of finding an exact solution for our problem
having modularity greater than two. Instead, we develop a more scalable, greedy approach.

\vspace{-2mm}
\subsection{Greedy Solution}
\vspace{-1mm}
We discuss our algorithm in two phases.

\vspace{1mm}
\subsubsection{Range Ratio Computation}
\label{sec:ratioComputation}
\vspace{-1mm}
Our first problem is: Given a total length $h$ and a specific way of combining the modules,
how shall we find the optimal range of all hash functions that correspond to different (combined) parts of the key?
In particular, consider $n$ modules: $\{x^{(1)},x^{(2)},\ldots,x^{(n)}\}$, and a specific way of combining them,
i.e., $\{y^{(1)},y^{(2)},\ldots,y^{(m)}\}$, where each $y^{(.)}$ is a singleton, or a combination of some
modules that are hashed together. For $k\ne j$, $y^{(k)}$ and $y^{(j)}$ do not have a common module.
Moreover, $y^{(k)}$ and $y^{(j)}$ are separately hashed for all $k \ne j$. Clearly, $m\leq n$, and each key is hashed
using $m$ independent hash functions. Given a length parameter $h$, we aim at finding the optimal ranges $a_1, a_2,\ldots, a_m$ of these hash functions,
where $a_1 \times a_2 \times \ldots \times a_m =h$.
\begin{figure}[t!]
\vspace{1mm}
\centering
\subfigure[]  {
    \includegraphics[scale=0.22]{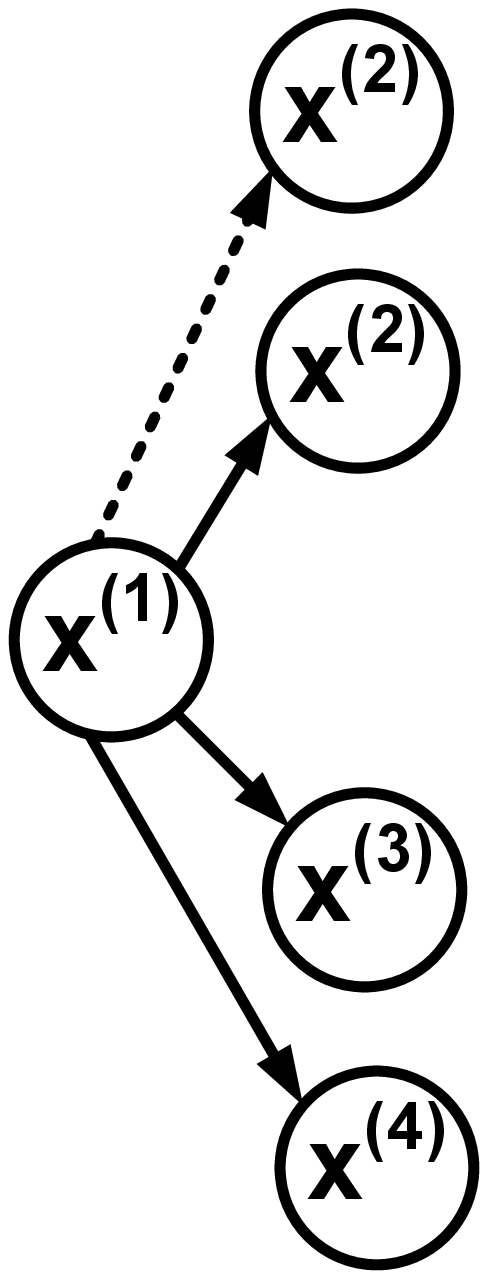}
    \label{fig:globaltree_1}
    }
    \quad
\subfigure[]  {
    \includegraphics[scale=0.22]{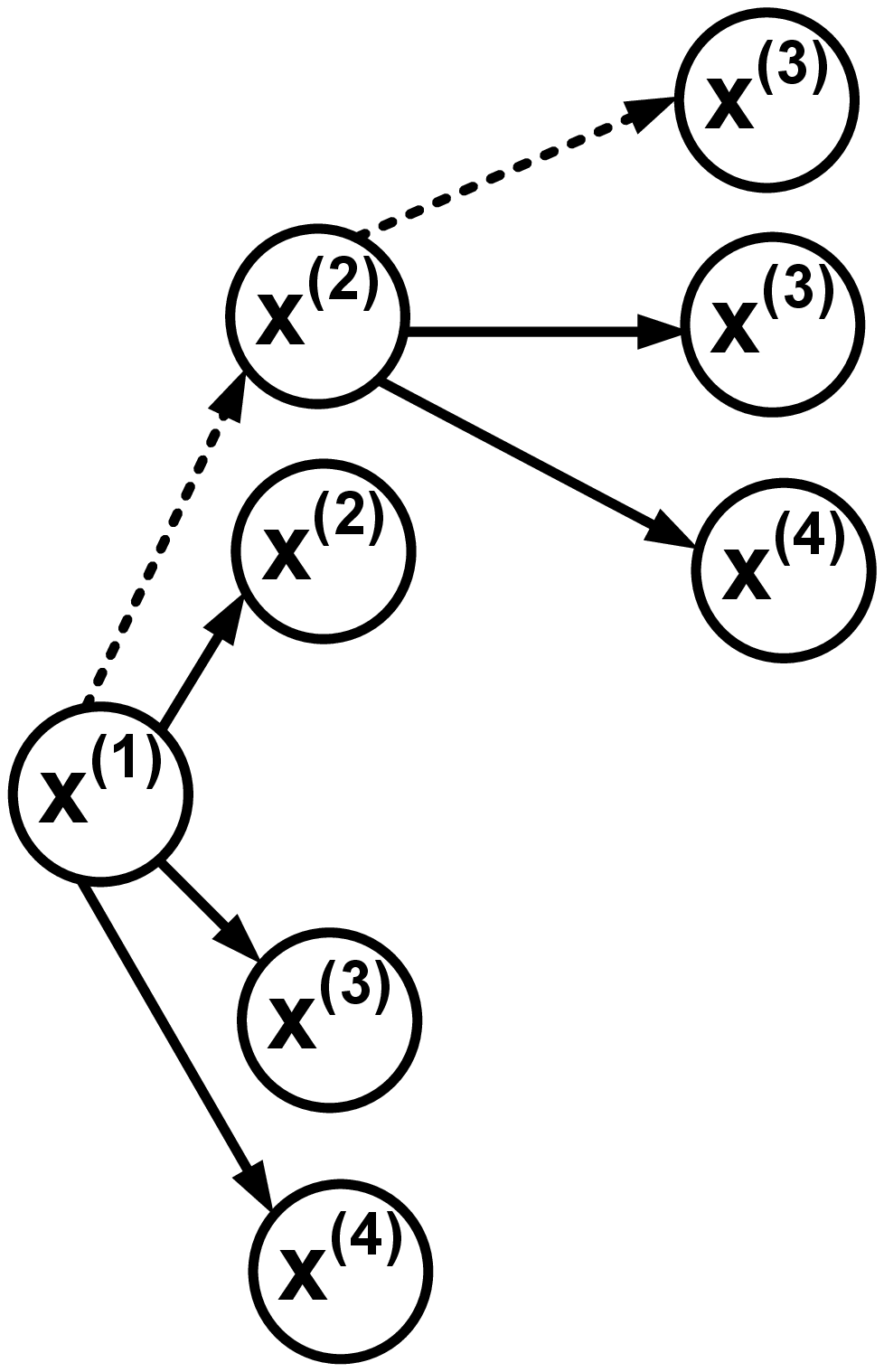}
    \label{fig:globaltree_2}
    }
     \quad
\subfigure[]  {
    \includegraphics[scale=0.22]{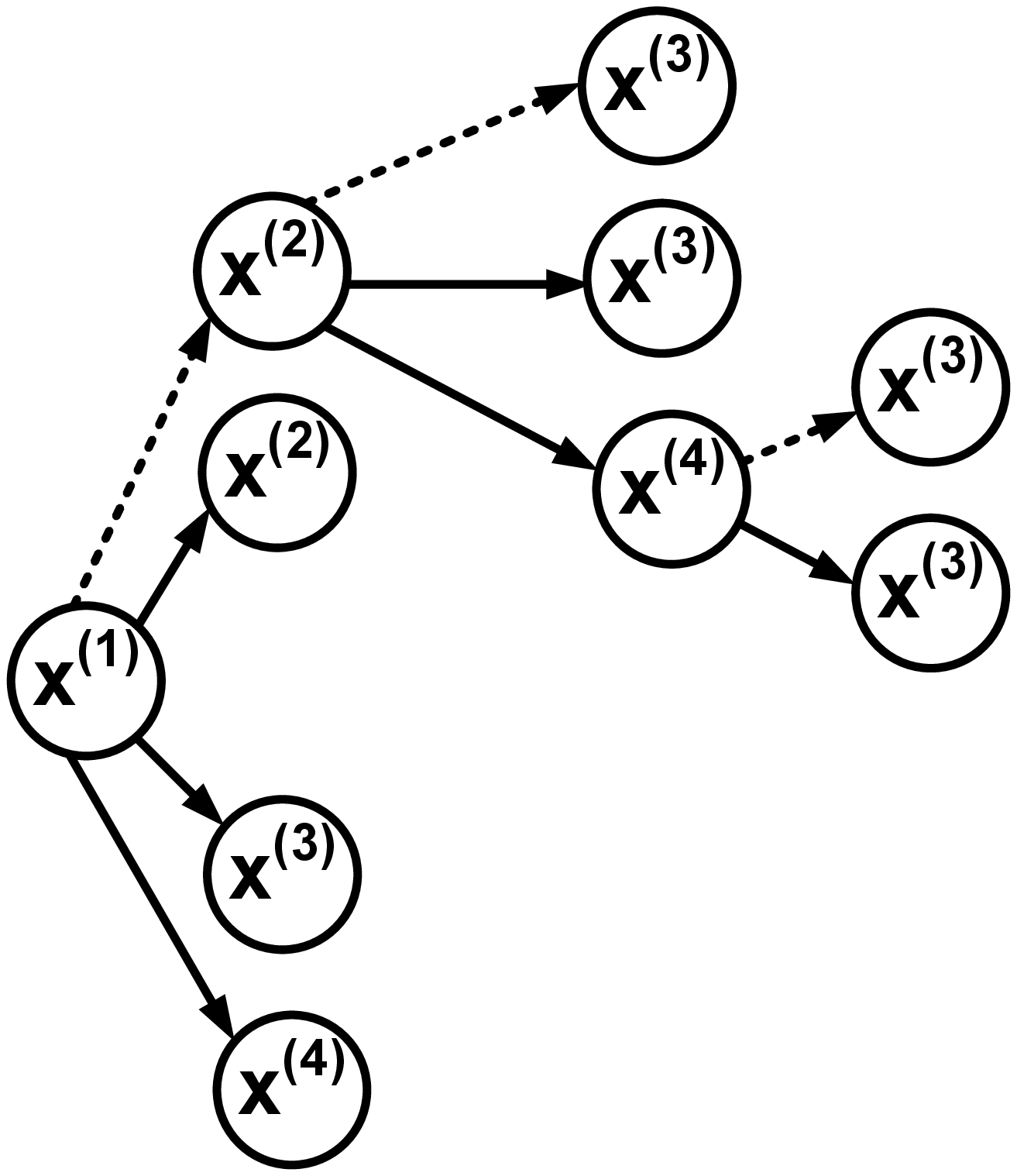}
    \label{fig:globaltree_3}
    }
\vspace{-3mm}
\caption{\small Greedy approach to find a good quality hashing strategy for keys with modularity $>2$.}
\label{fig:greedyTree}
\vspace{-6mm}
\end{figure}

Recall that for keys having modularity two, we developed our solution in \S~\ref{sec:opt_ab}. We now generalize
our method for keys with modularity more than two with a recursive strategy as follows.
We start by measuring the range ratio between $a_m$ (corresponding to the last hash function) and the rest.
Formally, we take the parts $y^{(1)},y^{(2)},\ldots,y^{(m-1)}$ as a combined part,
then we calculate the ratio of ranges $\beta_{m}=a_{m}/a_{1,2,\ldots,m-1}$, where $a_{m} \times a_{1,2,\ldots,m-1} = h$.
We compute $\beta_{m}$ with our method in \S~\ref{sec:opt_ab}, that is,
$\beta_{m}=\frac{1}{\alpha_{m}}$. We measure $\alpha_{m}=O(*,*,\ldots,*,y^{(m)})/O(y^{(1)}$, $y^{(2)},\ldots,y^{(m-1)},*)$.

Next, we recursively calculate the ratio $\beta_{m-1}=a_{m-1}/a_{1,2,\ldots,m-2}$ where $a_{m-1} \times a_{1,2,\ldots,m-2} = a_{1,2,\ldots,m-1}$.
We repeat this process until we find the ratio $\beta_{2}=a_2/a_1$.

Our recursive steps are illustrated in Figure~\ref{fig:ratioModel}.
The time complexity to compute the optimal values of $a_1,a_2,\ldots,a_m$ by our method is $\bigO(m)$, where $m\leq n$,
$n$ being the number of modules. We next discuss
the details of the greedy approach.

\vspace{1mm}
\subsubsection{Greedy Algorithm}
\vspace{-1mm}
Our greedy algorithm works by traversing the search space in a depth-first manner until a path of length $n$ has been found covering
all the $n$ modules. In Figure~\ref{fig:greedyTree}, we demonstrate our method with keys of modularity $4$:
$\{x^{(1)},x^{(2)},x^{(3)},x^{(4)}\}$. We start with the first module in order, which is $x^{(1)}$, and verify
four possibilities, that is, hashing $x^{(1)}$ separately (denoted by the dotted arrow between $x^{(1)}$ and $x^{(2)}$),
or combining $x^{(1)}$ with one of the remaining modules, $x^{(2)}$, $x^{(3)}$, or $x^{(4)}$ (denoted by solid arrows
between these nodes). Among these four possibilities, we greedily find the best strategy, e.g., in Figure~\ref{fig:globaltree_2}
we already selected the best strategy as to hash $x^{(1)}$ separately.

By traversing the dotted edge from $x^{(1)}$ to $x^{(2)}$, we arrive at the current module $x^{(2)}$. At this intermediate stage,
we have three choices: Hashing $x^{(2)}$ separately (denoted by the dotted arrow between $x^{(2)}$ and $x^{(3)}$),
or combining $x^{(2)}$ with one of the remaining modules, $x^{(3)}$ or $x^{(4)}$ (denoted by solid arrows
between these nodes). Again, we greedily find the best strategy, e.g., in Figure~\ref{fig:globaltree_3}
we already selected the best strategy as to combine $x^{(2)}$ and $x^{(4)}$.
\begin{algorithm}[tb!]
\caption{\small Greedily find optimal hashing of keys w/ modularity $\geq 2$}
\label{alg:greedy}
\small
\begin{algorithmic}[1]
\Require initial stream sample with keys having modularity $n\geq 2$, sketch parameters $h, w$
\Ensure optimal {\sf MOD-Sketch} configuration
\State $currentConfig$ $\rightarrow$ first module
\State $k \rightarrow 1$
\While{$k\leq n$}
\For{each $(n-k+1)$ hashing choices}
\State $m \rightarrow$ \#separate parts in the current choice; $m\leq k+1$
\State find optimal ranges $\{y^{(1)},y^{(2)},\ldots,y^{(m)}\}$ such that \newline $y^{(1)}\times y^{(2)}\times\ldots\times y^{(m)}=(h)^{\frac{k+1}{n}}$ (via Section~\ref{sec:ratioComputation})
\EndFor
\State $CurrentConfig$ $\rightarrow$ find the best hashing option among $(n-k+1)$ choices (via Section~\ref{sec:opt_mod_cm})
\State $k \rightarrow k+1$
\EndWhile
\State return $CurrentConfig$
\end{algorithmic}
\vspace{-1mm}
\end{algorithm}

In particular, at any intermediate stage $k$ ($1 \leq k \leq n$), we have computed the configuration
with $k$ modules, and next we have to select from $(n-k+1)$ choices in a greedy manner. For example in Figure~\ref{fig:globaltree_3},
we decided the configuration among three modules $x^{(1)}$, $x^{(2)}$, $x^{(4)}$, i.e., to keep $x^{(1)}$ separate and to combine
$x^{(2)}$, $x^{(4)}$. Now, we have two choices: Further combine $x^{(4)}$ with $x^{(3)}$, or keep them separate. We make a greedy
selection between these two choices, and finally terminate our method.

Our complete procedure is given in Algorithm~\ref{alg:greedy}. Note that by greedy selection, we only consider $\sum_{k=1}^n (n-k+1)=\bigO(n^2)$ choices
for combining the modules, as opposed to the exact number of ways $T(n)$ in which the modules can be combined. We have shown
in Table~\ref{tab:ways} that $T(n)$ increases at a higher rate compared to $2^n$, and this demonstrates the scalability of
our algorithm. Moreover, at any intermediate stage $k$, since we do not change the optimal configuration with $k$ modules
obtained previously, some of the optimal range ratio estimations from earlier stages could be re-used.
For example, assume in Figure~\ref{fig:globaltree_3}, out of two hashing options, the best one selected is to separate $x^{(4)}$ and $x^{(3)}$.
Thus, the three separate parts of the key are $x^{(1)}$, $x^{(2)}x^{(4)}$, and $x^{(3)}$. To find the optimal hashing ranges due to these three parts by following our recursive
method in Section~\ref{sec:ratioComputation}, one needs to compute the range ratio of the parts $x^{(1)}$ and $x^{(2)}x^{(4)}$. However, this ratio has been already computed
in Figure~\ref{fig:globaltree_2}, and therefore, can be re-used. Such re-using of range ratio estimation further improves our efficiency. 
\vspace{-2mm}
\section{Experimental Results}
\label{sec:experiments}
\vspace{-2mm}
\subsection{Environment Setup}
\subsubsection{Datasets}
\vspace{-2mm}
We consider six real-world stream datasets from two sources.
{\bf (1)} \emph{Twitter Communication Stream}: We obtain the {\em Twitter} graph dataset from
{\bf https://snap.} {\bf stanford.edu/}, which consists of all public tweets during a 7-month period from June 1, 2009
to December 31, 2009. Each edge (directed) in our dataset represents a communication between two users
in the form of a re-tweet. The edge frequency is defined as the number of communications between the corresponding
source and target users. Clearly, every item (i.e., an edge) in this dataset has modularity two, consisting of
a source and a target node.
{\bf (2)} \emph{IPv4 Trace Stream}: We use two IP trace streams from the IPv4 Routed /24 Topology dataset
({\bf http://www.caida.org/data/overview/}),
which contains the raw IPv4 team-probing data from January 1, 2008 to December 31, 2008 by CAIDA.
We obtain two datasets of IP address streams from source-to-destination and source-to-respond traces,
and refer to them as {\em IPv4-1} and {\em IPv4-2}, respectively. Each item in both datasets contains two 4-bytes IP addresses.
Therefore, both {\em IPv4-1} and {\em IPv4-2} have modularity eight, denoted as {\em IPv4-1\#8} and {\em IPv4-2\#8}.
Next, we generate modularity-two datasets from {\em IPv4-1\#8} and {\em IPv4-2\#8} by hashing each IP address to one integer id,
thereby obtaining datasets {\em IPv4-1\#2} and {\em IPv4-2\#2}. Similarly, we generate modularity-four datasets by hashing each IP address with two integer ids
 --- the first integer corresponds to the first 2-bytes of the IP address, and the second one corresponds to the next 2-bytes of the IP address. In this way,
 we obtain another two datasets {\em IPv4-1\#4} and {\em IPv4-2\#4}, having modularity four.

In Table~\ref{tab:dataset}, the flat stream size represents the total stream size that contains repetition of items.
The compressed stream size, in contrast, is defined as the size of all distinct items that have nonzero frequency, along with their frequency counts.
Both flat stream and compressed stream can answer our queries with complete accuracy. {\em We shall demonstrate in our experiments that {\sf MOD-Sketch}, though a
small fraction of the flat and compressed stream representations, achieves reasonably high accuracy}. 

In Table~\ref{tab:twopart}, we show additional
statistics for our three datasets with modularity two.
\begin{table} [tb!]
\centering
\scriptsize
\vspace{-1mm}
\caption{\small Data stream characteristics and sizes}
\label{tab:dataset}
\vspace{-2mm}
\begin{tabular}{lc|c|cc|cc}
{\textsf Datasets} & {\textsf Modul} & {\textsf \# Distinct} & {\textsf Agg. item} & {\textsf Max. item} & {\textsf Flat} & {\textsf Compressed}\\
& {\textsf -arity} & {\textsf items} & {\textsf frequency}   & {\textsf frequency} & {\textsf stream} & {\textsf stream} \\
&                  &                 &                       &                     & {\textsf(GB)}    & {\textsf(GB)}    \\ \hline \hline
\emph{Twitter}         & 2  & 78\,508\,963  & 151$\times 10^{6}$    &   17\,149      &   3.69    &          0.15              \\ \hline \hline
\emph{IP-1$\#2$}     & 2  & 94\,820\,182  & 6\,204 $\times 10^{6}$&   123\,614     &   7.23    &          1.72              \\ \hline
\emph{IP-1$\#4$}     & 4  & 94\,820\,182  & 6\,204 $\times 10^{6}$&   123\,614     &   9.50    &          2.26              \\ \hline
\emph{IP-1$\#8$}     & 8  & 94\,820\,182  & 6\,204 $\times 10^{6}$&   123\,614     &   13.42   &          4.19              \\ \hline \hline
\emph{IP-2$\#2$ }    & 2  & 94\,890\,903  & 6\,562 $\times 10^{6}$&   106\,956     &   17.98   &          1.81              \\ \hline
\emph{IP-2$\#4$}     & 4  & 94\,890\,903  & 6\,562 $\times 10^{6}$&   106\,956     &   22.79   &          3.29              \\ \hline
\emph{IP-2$\#8$}     & 8  & 94\,890\,903  & 6\,562 $\times 10^{6}$&   106\,956     &   30.35   &          5.05              \\
\end{tabular}
\vspace{-3mm}
\end{table}
\begin{table} [tb!]
\centering
\scriptsize
\vspace{-1mm}
\caption{\small Additional statistics for streams having modularity two}
\label{tab:twopart}
\vspace{-2mm}
\begin{tabular}{l||cc}
{\textsf Datasets} & {\textsf\#Source nodes} & {\textsf \#Target nodes} \\ \hline \hline
\emph{Twitter}     &  4\,790\,726      & 15\,062\,341 \\ \hline
\emph{IP-1$\#$2} &  7\,234\,121      & 665\,279 \\ \hline
\emph{IP-2$\#$2} &  8\,352\,656      & 697\,121  \\
\end{tabular}
\vspace{-6mm}
\end{table}
\vspace{1mm}
\subsubsection{Comparing methods}
\vspace{-1mm}
We compare {\sf MOD-Sketch} with three following methods.
{\bf (1)} {\sf Count-Min} \cite{countmin}, i.e., concatenate $n$ ordered modules of the key and hash it directly with a hash function of range $h$.
We use $w$ such pairwise independent hash functions.
{\bf (2)} {\sf Equal-Sketch}, i.e., separately hash $n$ modules with $n$ independent hash functions, each having range $(h)^{1/n}$.
We use $w$ such pairs of hash functions. In the context of graph edge stream (i.e., having modularity two), similar methods were
explored, e.g., {\sf TCM} \cite{TCM16} and {\sf gMatrix} \cite{KA17}.
{\bf (3)} {\sf Exhaustive}, i.e., empirically evaluate all $T(n)$ possibilities of combining the modules of a key (the exact value of $T(n)$ can be
obtained from Table~\ref{tab:ways}). For each possibility, experimentally find the best choice of hash function ranges
corresponding to different (combined) parts of the key. Finally, select the best option that minimizes the frequency estimation error. Clearly,
the {\sf Exhaustive} approach is very expensive, and it does not scale well beyond modularity $n>4$.

\vspace{1mm}
\subsubsection{System description}
\vspace{-1mm}
We implement our code
in Python, and perform experiments on a single core of 2.40GHz Xeon server. {\em Each experiment uses
less than 1GB of the main memory}, and our empirical results are averaged over 10 runs.
\begin{figure*}[t!]
\vspace{-2mm}
\centering
\subfigure[\small{\emph{Twitter}, h=$10^6$, w=10}]  {
    \includegraphics[scale=0.32]{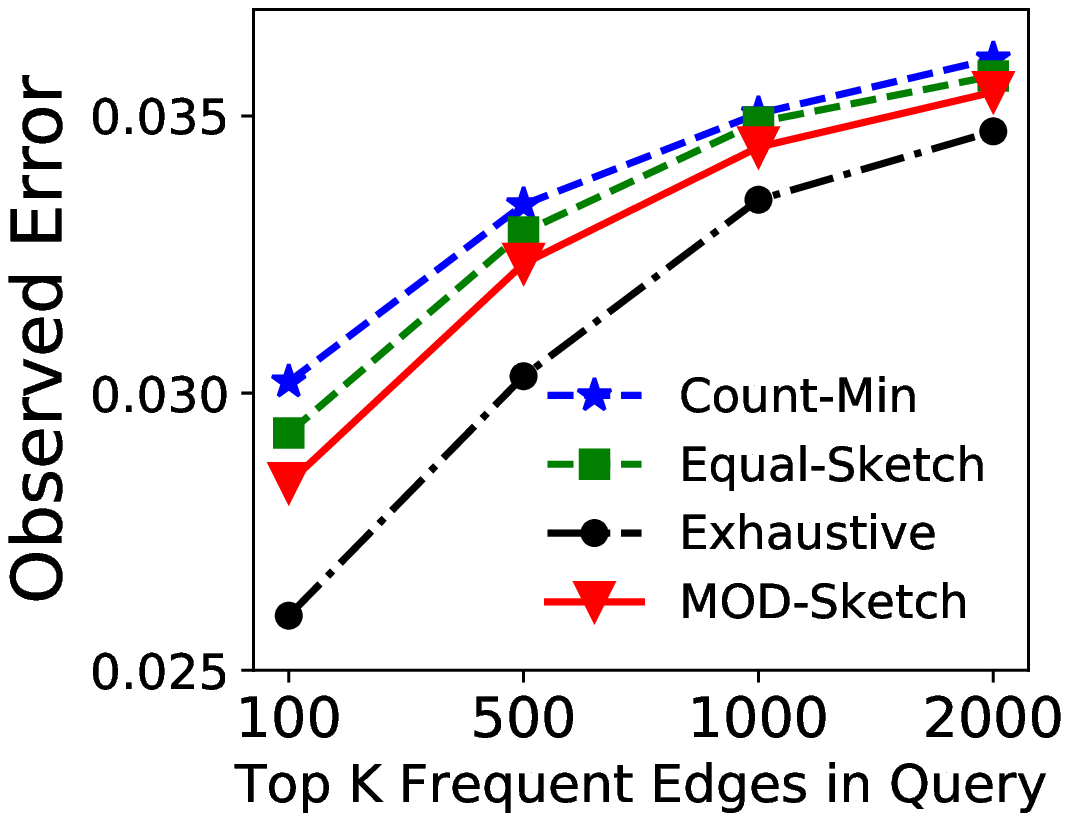}
    \label{fig:tweet-topk}
}
\subfigure[\small{\emph{IPv4-1$\#2$}, h=$4\times10^6$, w=10}]  {
    \includegraphics[scale=0.32]{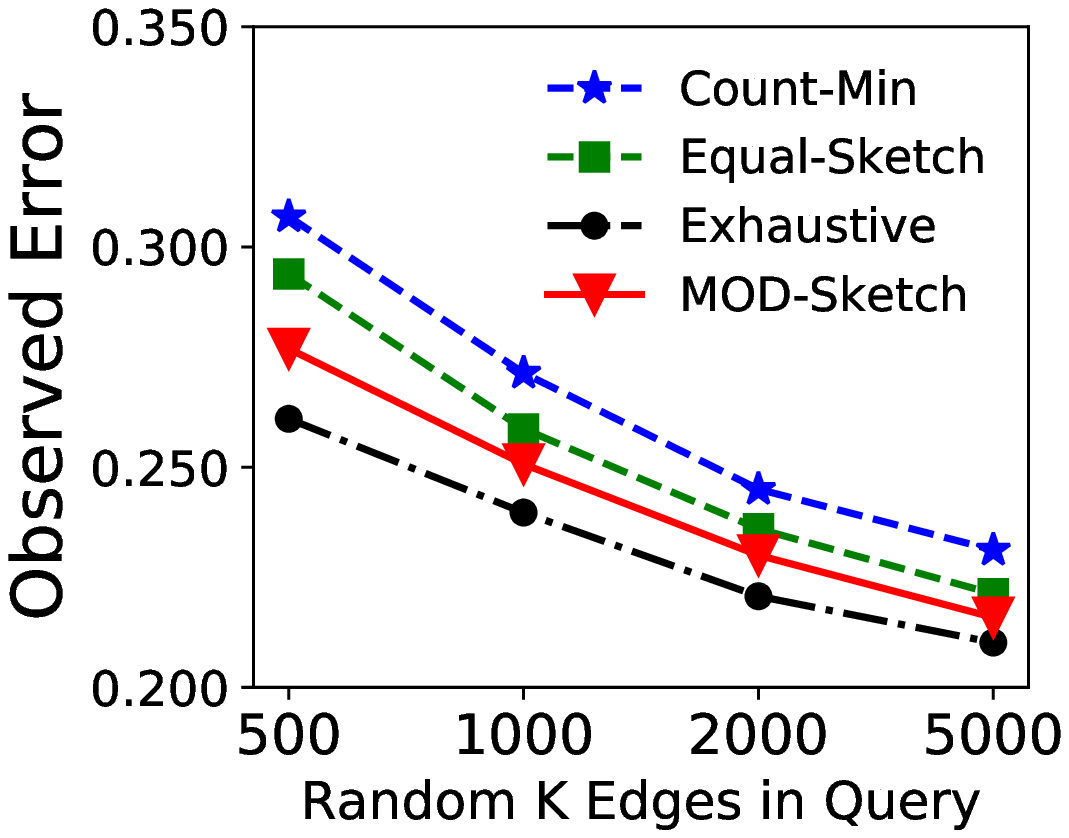}
    \label{fig:ipv41-randomk}
}
\subfigure[\small{\emph{IPv4-2$\#2$}, w=10, Top 500 Edges in Query}]  {
    \includegraphics[scale=0.32]{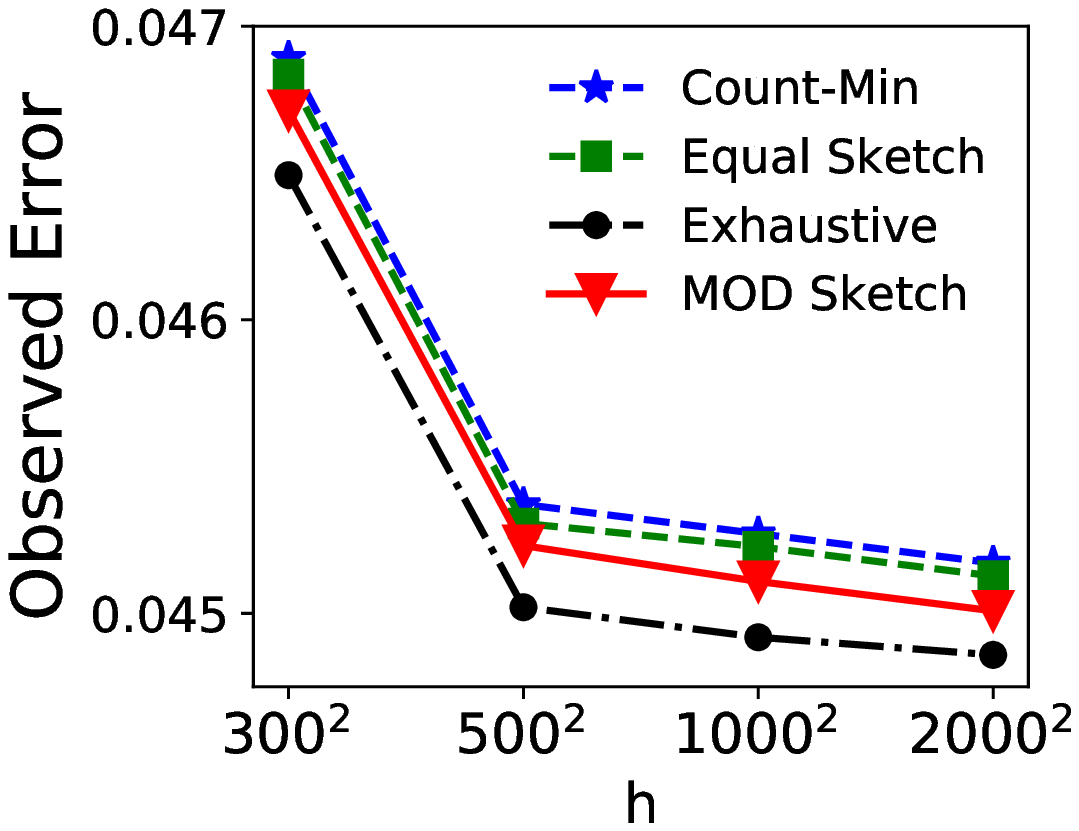}
    \label{fig:ipv42-top-h}
}
\vspace{-4mm}
\caption{\small Effectiveness of edge frequency estimation queries. Each sketch size=$7.2$MB, $20$MB, $80$MB, $320$MB when $h$=$9\times10^4$, $25\times10^4$, $10^6$, $4\times10^6$, respectively. {\sf MOD-Sketch} is generated by sampling 2\% of the stream.}
\label{fig:twopartoe}
\vspace{-2mm}
\end{figure*}
\begin{figure*}[t!]
\vspace{-2mm}
\centering
\subfigure[\small{\emph{Twitter}, h=$4\times10^6$, w=10, Rand. 1000 Edges in Query}]  {
    \includegraphics[scale=0.29]{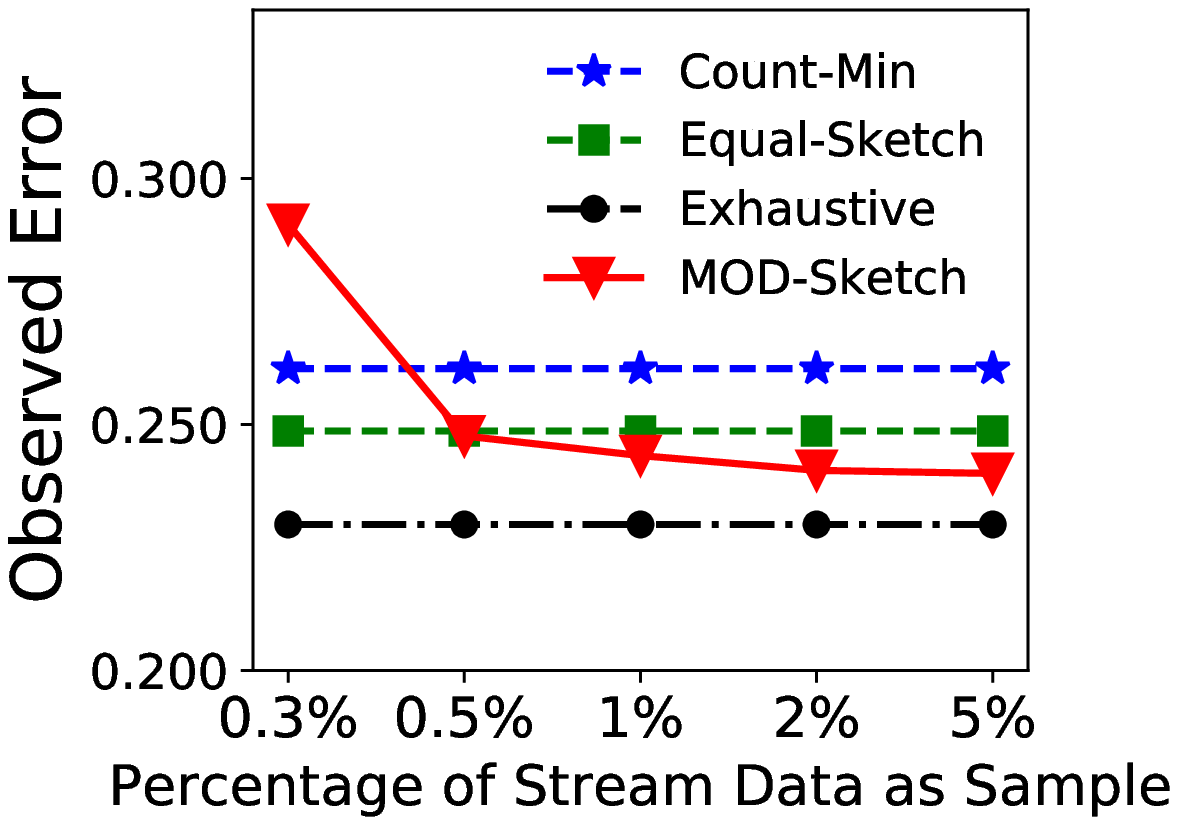}
    \label{fig:ipv41-random-sample}
}
\subfigure[\small{\emph{IPv4-1$\#2$}, h=$10^6$, w=10, Top 100 Edges in Query}]  {
    \includegraphics[scale=0.29]{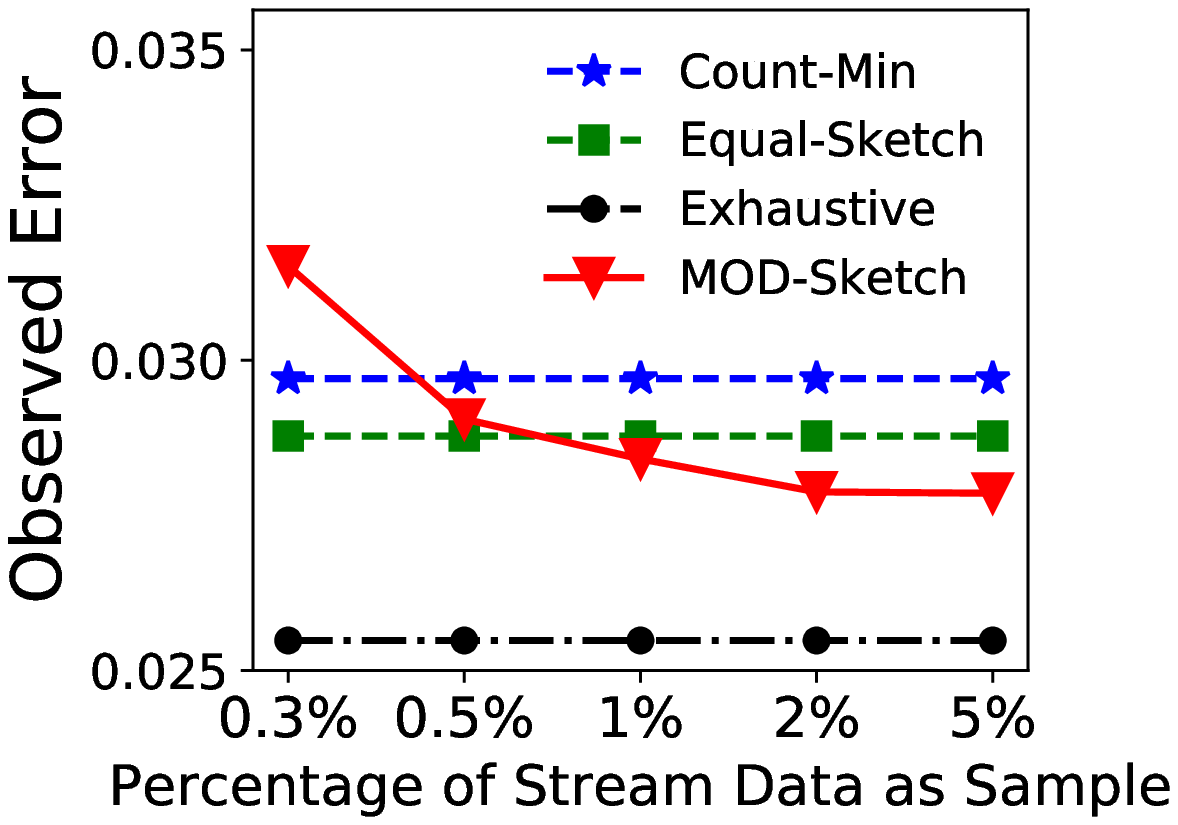}
    \label{fig:tweet-Sample-Size}
}
\subfigure[\small{\emph{IPv4-2$\#2$}, h=$10^6$, w=10, Top 500 Edges in Query}]  {
    \includegraphics[scale=0.29]{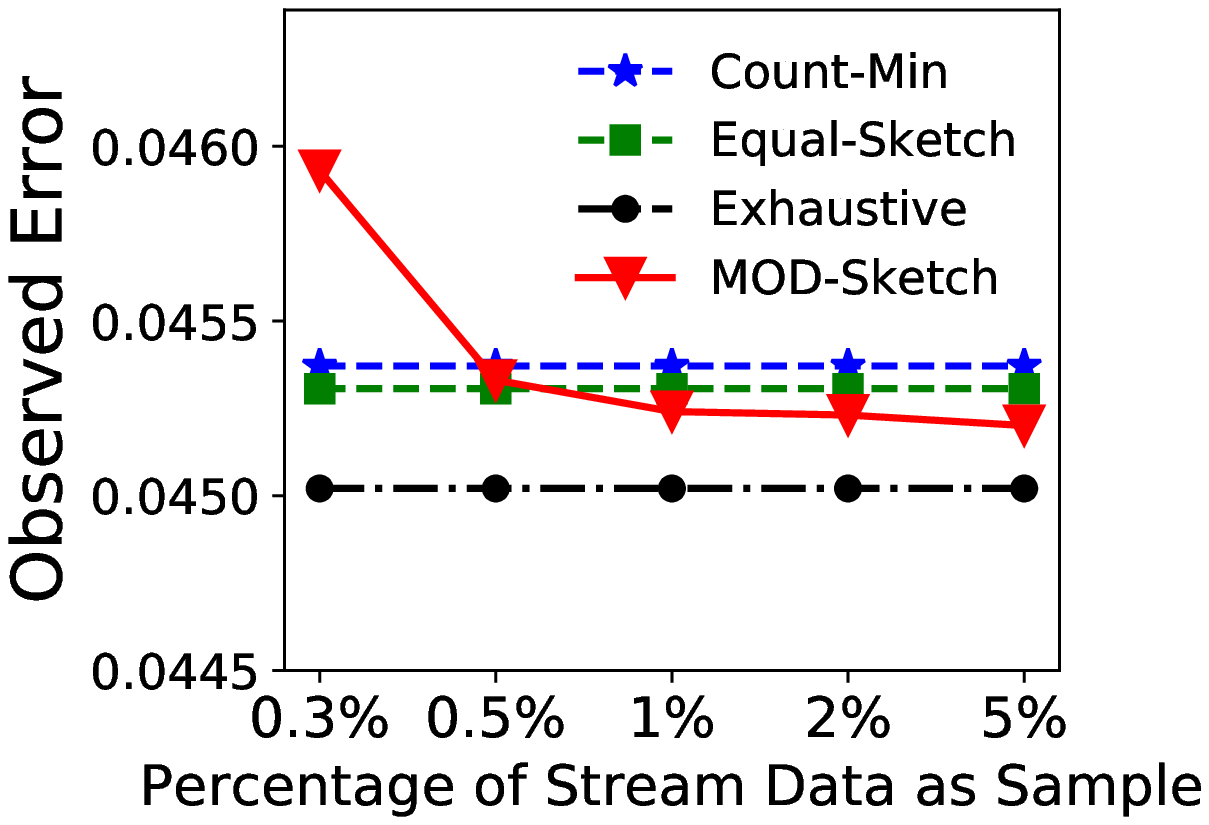}
    \label{fig:ipv42-top-sample}
}
\vspace{-4mm}
\caption{\small Effectiveness of edge frequency estimation queries using different size of data sample for {\sf MOD-Sketch}.
Each sketch size=$80$MB, $320$MB when $h$=$10^6$, $4\times10^6$, respectively.}
\label{fig:ipv41randomSample}
\vspace{-2mm}
\end{figure*}
\begin{figure*}[t!]
\vspace{-2mm}
\centering
\subfigure[\small{\emph{Twitter}}]  {
    \includegraphics[scale=0.31]{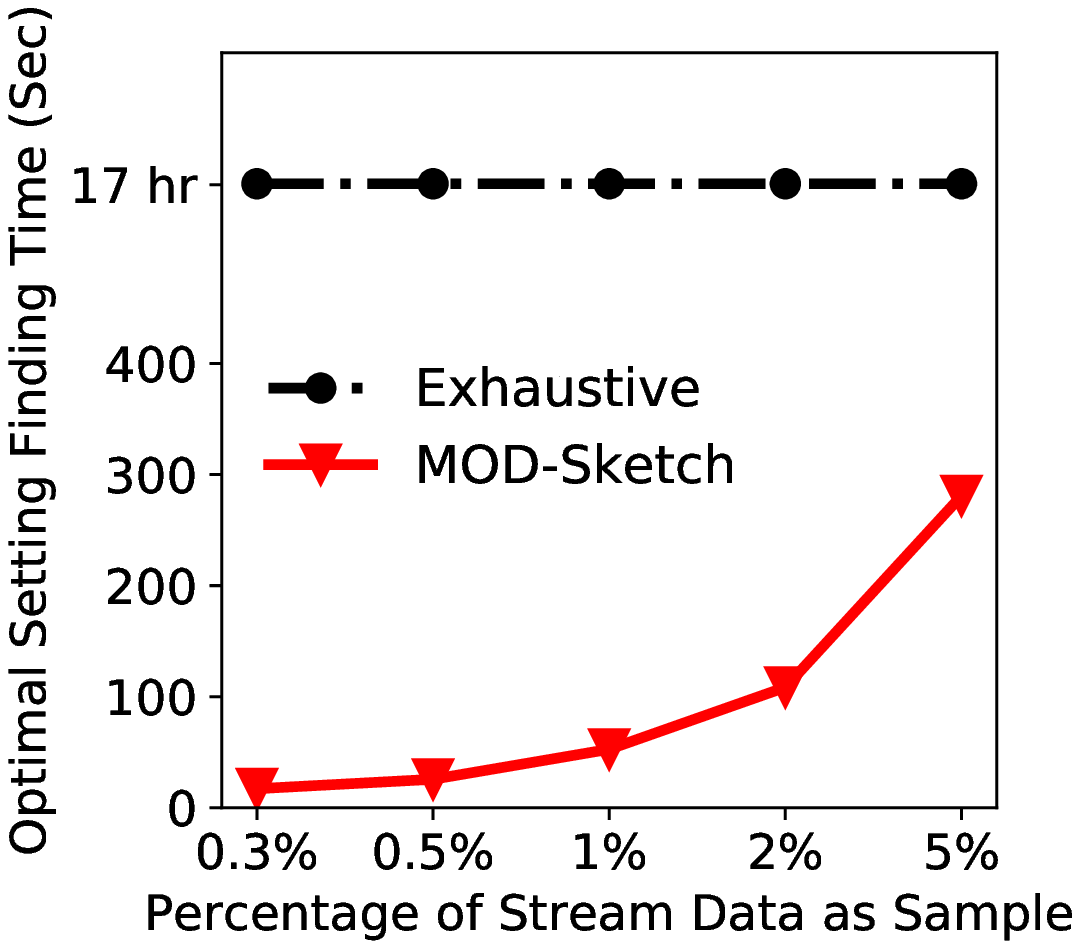}
    \label{fig:twitter-time-sample}
}
\subfigure[\small{\emph{IPv4-1$\#$2}}]  {
    \includegraphics[scale=0.31]{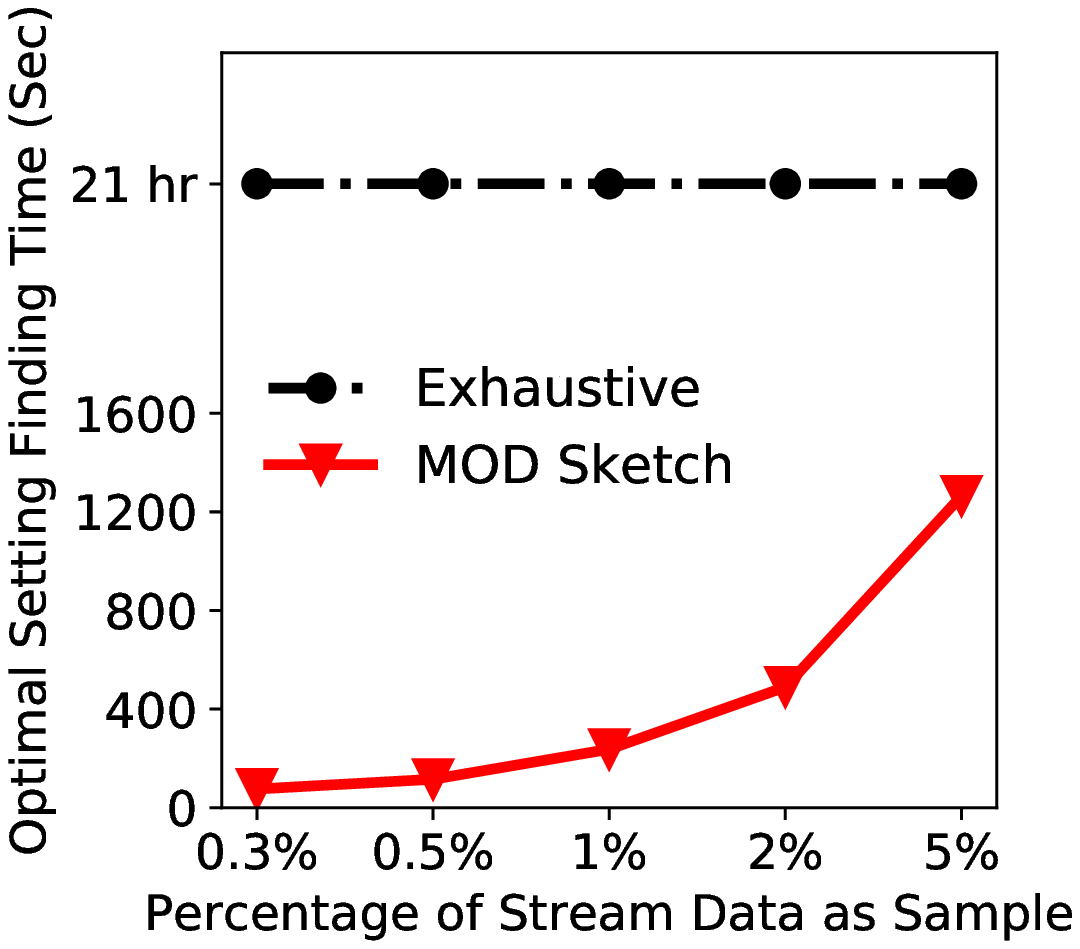}
    \label{fig:ipv41-time-sample}
}
\subfigure[\small{\emph{IPv4-2$\#$2}}]  {
    \includegraphics[scale=0.31]{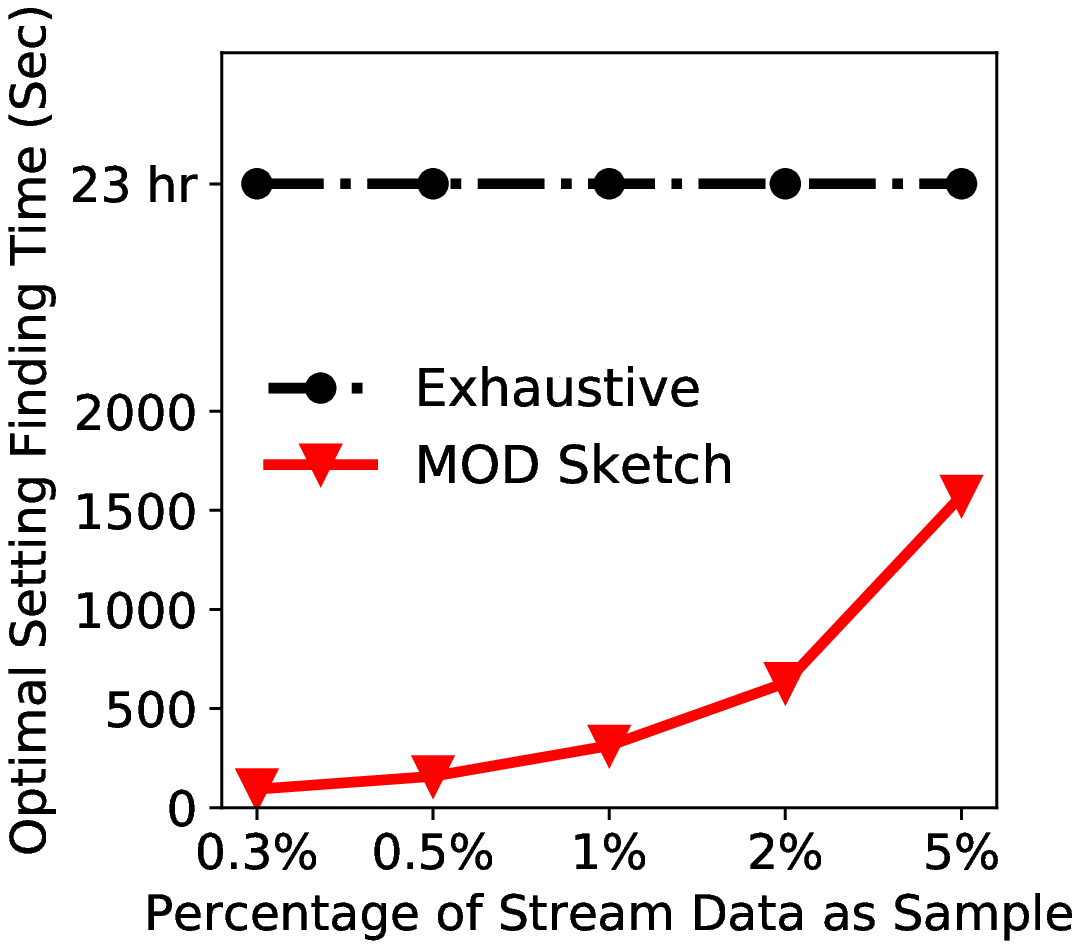}
    \label{fig:ipv42-time-sample}
}
\vspace{-4mm}
\caption{\small Efficiency of constructing {\sf MOD-Sketch} vs. obtaining the experimentally-best-sketch {\sf Exhaustive}.}
\label{fig:timepart2}
\vspace{-2mm}
\end{figure*}
\begin{figure*}[t!]
\vspace{-2mm}
\centering
\subfigure[\small{\emph{IPv4-1}, w=10}]  {
    \includegraphics[scale=0.32]{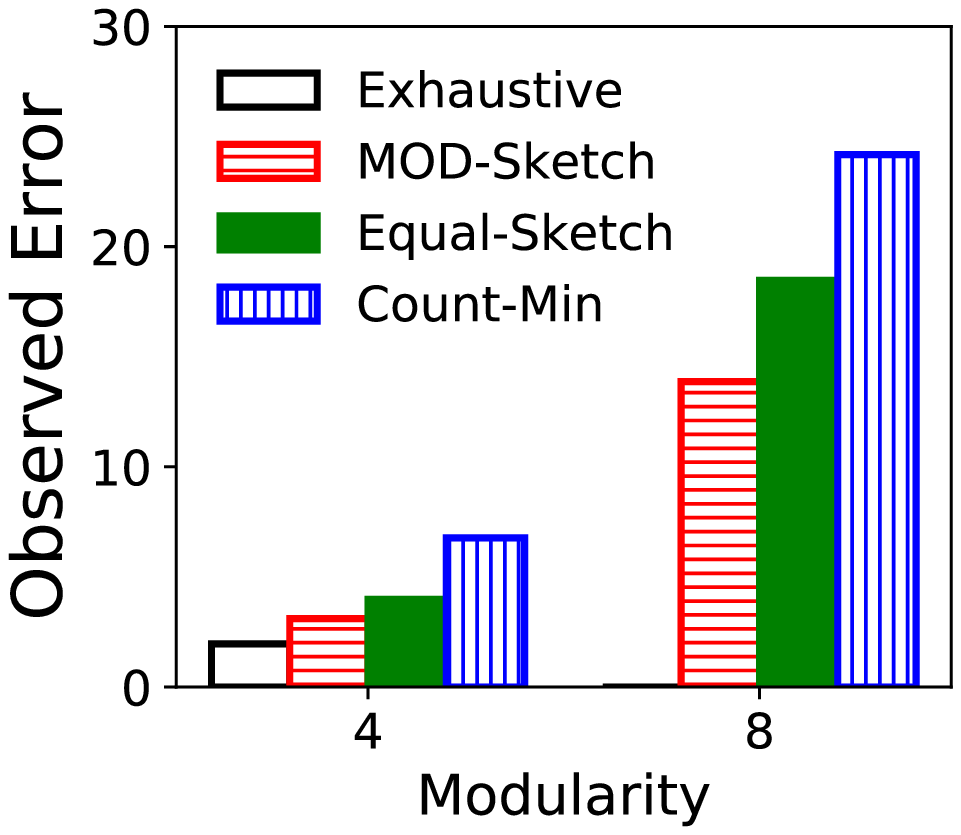}
    \label{fig:ipv41Top}
}
\subfigure[\small{\emph{IPv4-2}, w=10}]  {
    \includegraphics[scale=0.32]{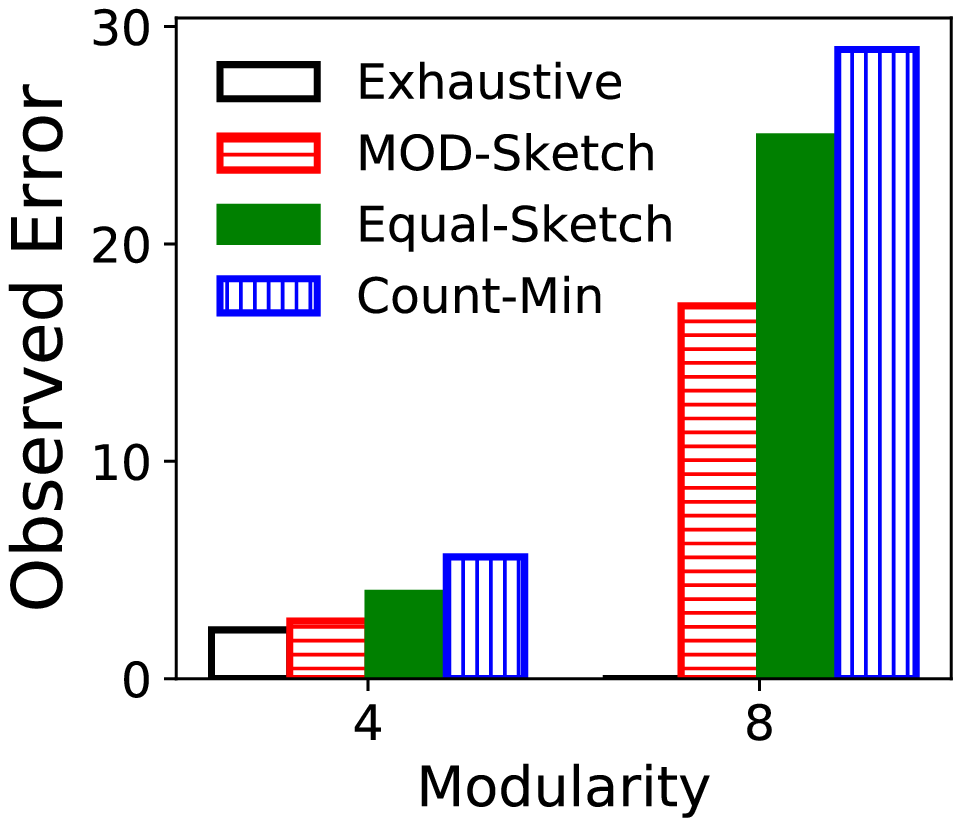}
    \label{fig:ipv42Rad}
}
\subfigure[\small{\emph{IPv4-1$\#$4}}]  {
    \includegraphics[scale=0.32]{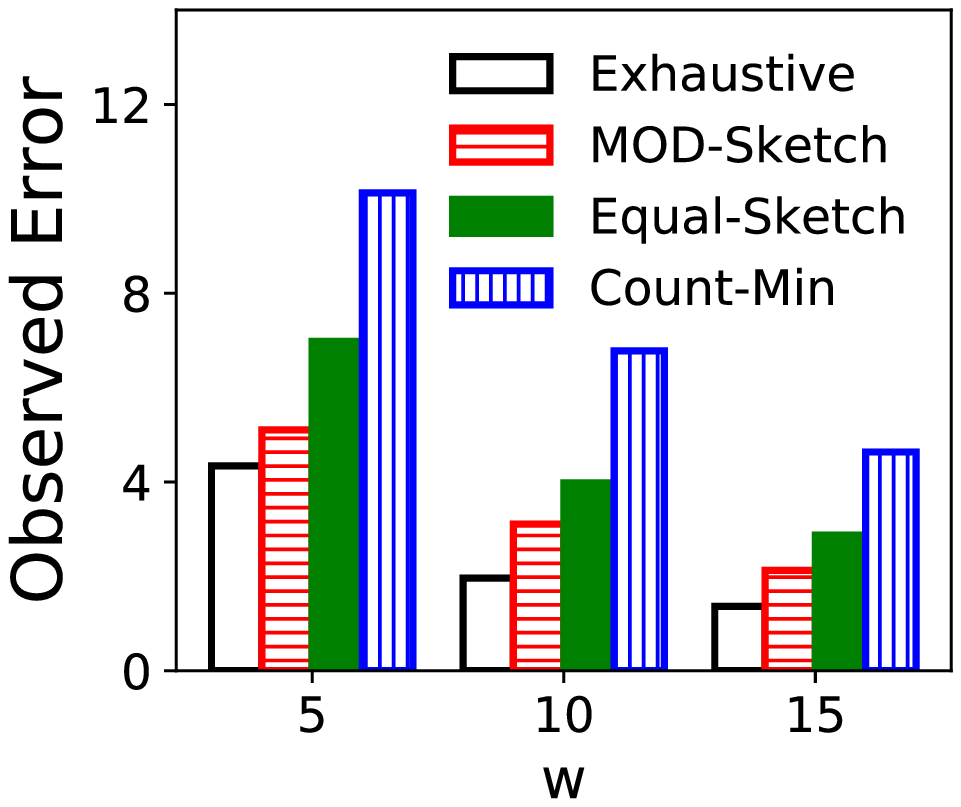}
    \label{fig:ipv41w}
}
\vspace{-4mm}
\caption{\small Effectiveness of edge frequency estimation queries for varied modularity of streaming data. The total range is set as $h= 4\times10^6$ in all three figures.  Top-100 query. Each sketch size=160MB, 320MB, 480MB when w=5, 10, 15, respectively. {\sf MOD-Sketch} is generated by sampling 2\% of the stream. For modularity 8, {\sf Exhaustive} does not finish in 100 hours.}
\label{fig:dimplot}
\vspace{-5mm}
\end{figure*}
\begin{figure*}[t!]
\centering
\subfigure[\small{\emph{IPv4-1}, w=10}]  {
    \includegraphics[scale=0.32]{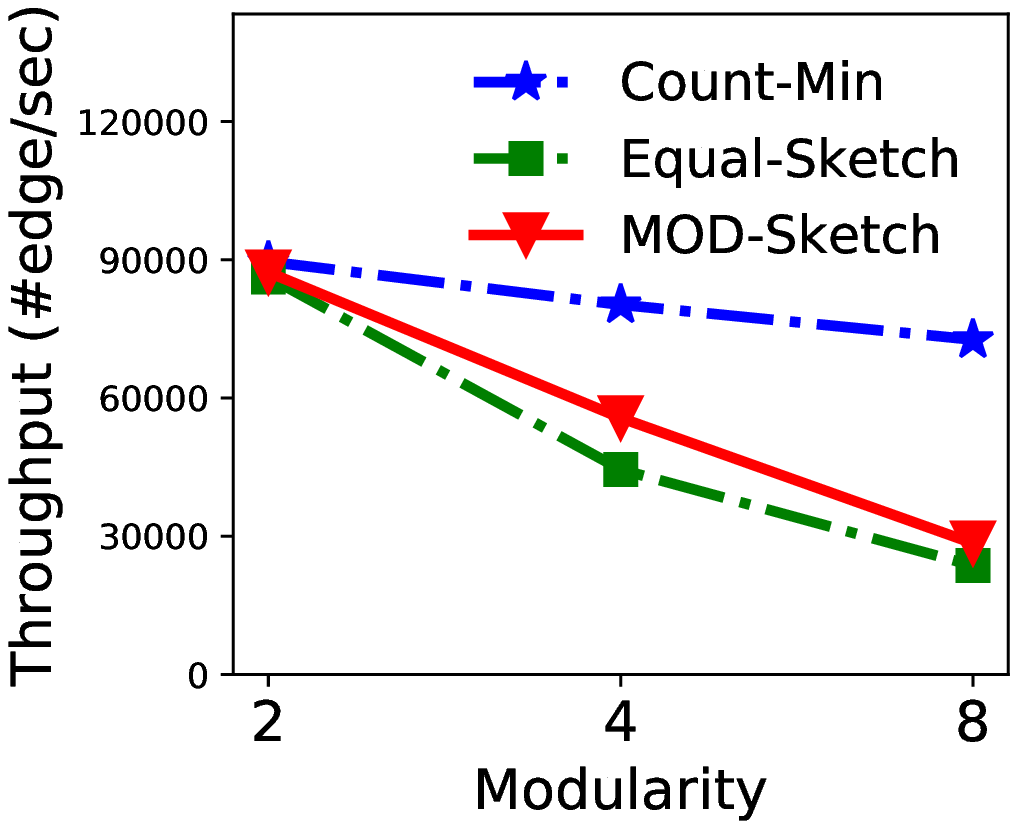}
    \label{fig:ThroughputIpv41dim}
}
\subfigure[\small{\emph{IPv4-2}, w=10}]  {
    \includegraphics[scale=0.32]{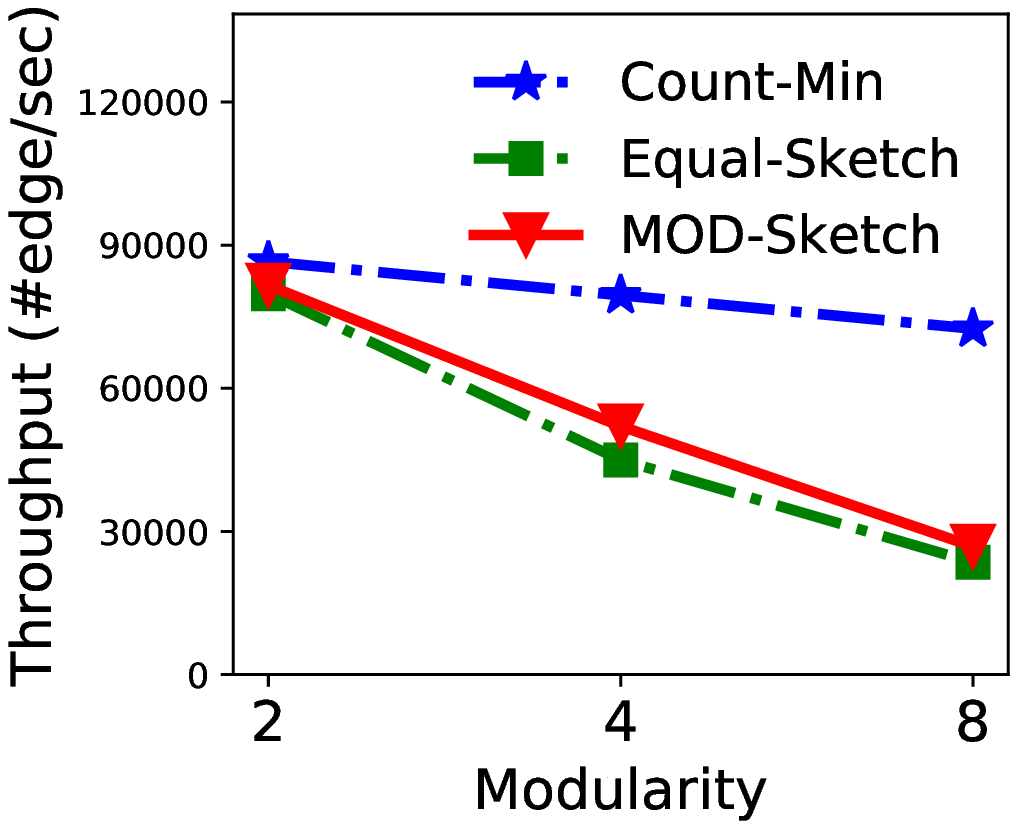}
    \label{fig:ThroughputIpv42dim}
}
\subfigure[\small{\emph{IPv4-1$\#4$}, dim=4}]{
    \includegraphics[scale=0.32]{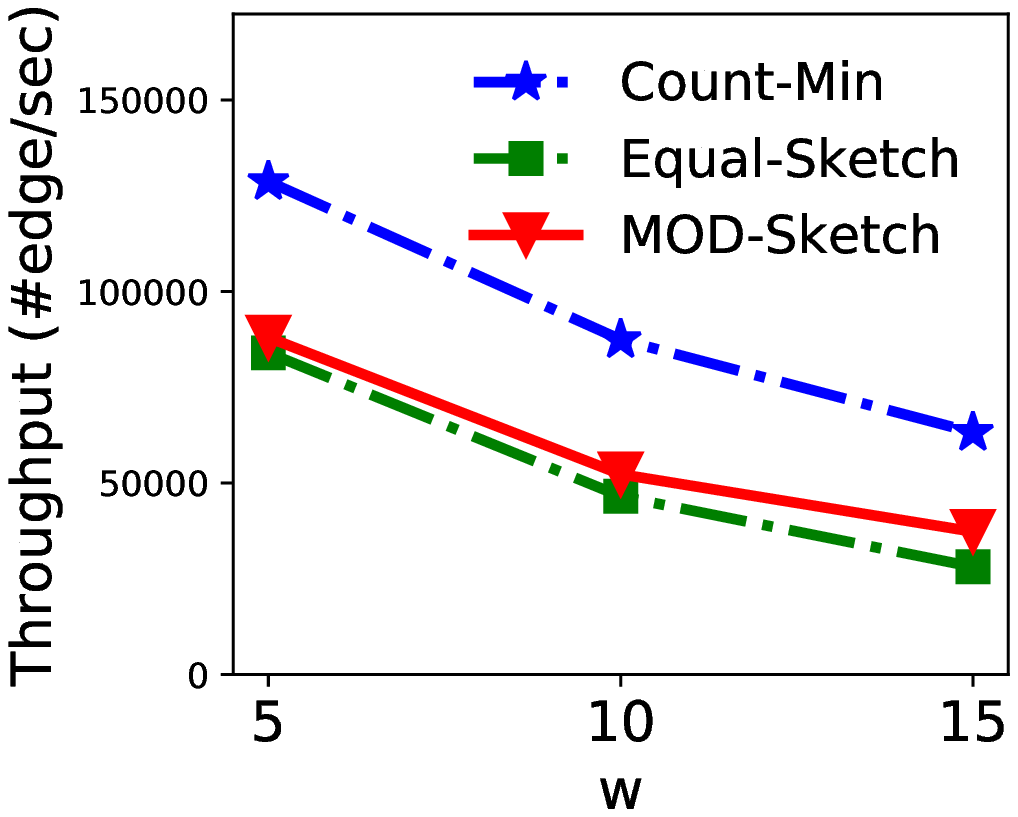}
    \label{fig:ThroughputIpv4W}
}
\vspace{-4mm}
\caption{\small Throughput of stream processing. The total range is set as $h= 4\times10^6$ in all three figures.}
\label{fig:throughput}
\vspace{-6mm}
\end{figure*}
\vspace{-1mm}
\subsubsection{Evaluation metrics}
\vspace{-1mm}
We present the accuracy of frequency estimation queries using observed error \cite{countmin},
which is measured as the difference between the estimated frequency and the true frequency,
accumulated over all the items queried. 
The observed error is expressed as a ratio over the aggregate true frequency of all the items queried.
\vspace{-2mm}
\begin{equation}
\text{Observed error} = \displaystyle \frac{\sum_{i \in Query}{|\text{estimated freq}_{(i)}-\text{true freq}_{(i)}|}}{\sum_{i \in Query}{\text{true freq}_{(i)}}} \nonumber
\vspace{-1mm}
\end{equation}
We generate two types of query sets: {\bf (1)} {\em Top-$k$ query} consists of the top-$k$ high frequency items. {\bf (2)} {\em Random-$k$ query} consists
of $k$ randomly selected items from the data stream.
\vspace{-2mm}
\subsection{Results for Modularity Two}
\vspace{-1mm}
In Figures~\ref{fig:twopartoe} and \ref{fig:ipv41randomSample}, we show the accuracy of edge frequency queries over
our datasets with modularity two, under varying $k$ of both top-$k$ and random-$k$ queries, and for different range parameter $h$.
We notice that in all experiments, {\sf MOD-Sketch} obtains smaller observed error compared to {\sf Count-Min} and {\sf Equal-Sketch}.
As an example, in \emph{Twitter}, {\sf MOD-Sketch} with $a=434$ and $b=2304$ has a relatively lower observed error,
compared to {\sf Count-Min} of one combined range $h=10^6$ and {\sf Equal-Sketch} of two equal ranges $a=b=10^3$.
Compared to {\sf Exhaustive}, which empirically finds the optimal parameters: $a=470$ and $b=2127$, the setting of {\sf MOD-Sketch} is very similar.
Further delving into the {\em Twitter} dataset, we find that the distinct number of target nodes is more than
that of source nodes. This justifies why our method {\sf MOD-Sketch}, as well as {\sf Exhaustive} report $b>a$ in the optimal setting.
We find similar observations with {\em IPv4-1$\#2$} and {\em IPv4-2$\#2$}.

While {\sf Exhaustive} produces the least observed error, finding the optimal parameters experimentally by considering
various combinations requires about 20 hours, which is not affordable (Figure~\ref{fig:timepart2}). In comparison,
{\sf MOD-Sketch} requires a few minutes (5$\sim$20 minutes) to find good-quality $a$ and $b$ values that result in
comparable accuracy with {\sf Exhaustive}. Moreover, the observed error of {\sf MOD-Sketch} reduces when the size of the data sample for
its parameter estimation increases. We find that with about 2\% of the stream, the observed error converges (Figure~\ref{fig:ipv41randomSample}), therefore in our experiments, we use 2\% of the stream for parameter estimation.

%
%
\vspace{-2mm}
\subsection{Results for Modularity $>$ Two}
\vspace{-1mm}
We analyze the accuracy of edge frequency estimation over our datasets with modularity 4 and 8, and
with varying $w$ (Figure~\ref{fig:dimplot}). The observed error increases with higher modularity.
However, the observed error of {\sf MOD-Sketch} is always smaller than that of {\sf Equal-Sketch} and {\sf Count-Min},
and is also comparable to {\sf Exhaustive}. In fact, for modularity 8, the observed error of {\sf MOD-Sketch}
is almost half of that due to {\sf Count-Min} and {\sf Equal-Sketch}.

\begin{figure}[t!]
\vspace{-2mm}
\centering
\subfigure[\emph{IPv4-1}]  {
    \includegraphics[scale=0.29]{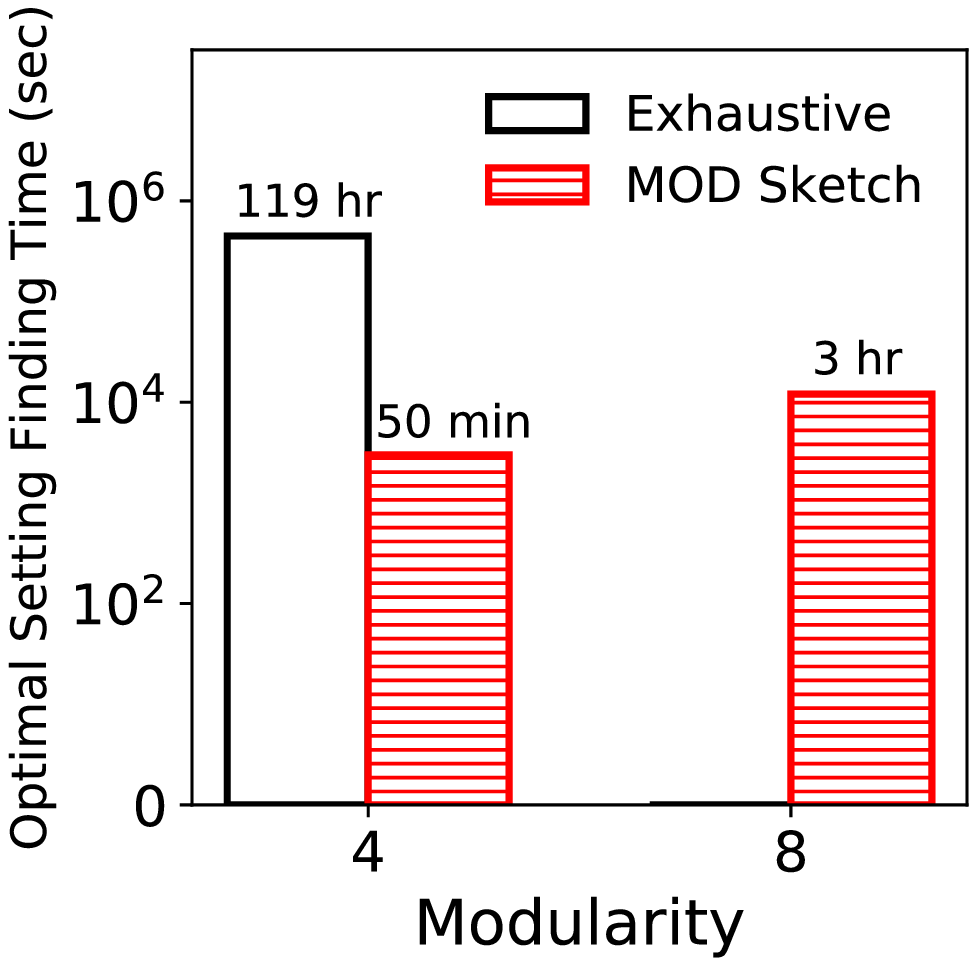}
    \label{fig:timeipv41dim}
}
\qquad
\subfigure[\emph{IPv4-2}]  {
    \includegraphics[scale=0.29]{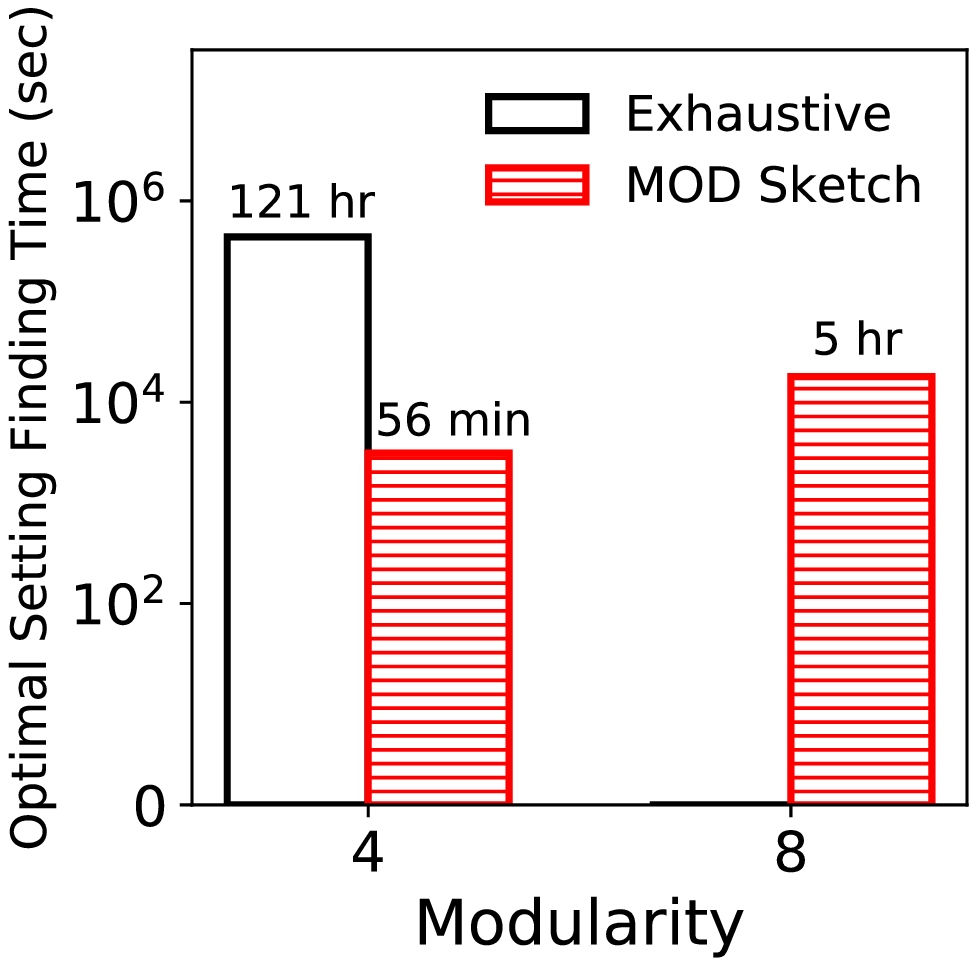}
    \label{fig:timeipv42dim}
}
\vspace{-4mm}
\caption{\small Efficiency of constructing {\sf MOD-Sketch} vs. obtaining the experimentally-best-sketch {\sf Exhaustive} in highly-modular stream data.
For modularity 8, {\sf Exhaustive} does not finish in 100 hours.}
\label{fig:timedim}
\vspace{-4mm}
\end{figure}

Moreover, the running time to find the optimal parameters
using {\sf Exhaustive} is at least two orders of magnitude higher than that of {\sf MOD-Sketch} for
modularity 4. In fact, with modularity 8, {\sf Exhaustive} does not complete within 100 hours (Figure~\ref{fig:timedim}).
{\em These results demonstrate the efficiency and scalability of {\sf MOD-Sketch} over highly-modular data streams.}
%
%
\vspace{-6mm}
\subsection{Stream Processing Throughput}
\vspace{-1mm}
The throughputs of {\sf MOD-Sketch}, {\sf Count-Min}, and {\sf Equal-Sketch} are
close when the modularity is small (Figure~\ref{fig:throughput}). For streams with higher modularity,
the throughput of {\sf MOD-Sketch} is lower than that of {\sf Count-Min},
but higher than that of {\sf Equal-Sketch}. This can be explained as follows. For an item with modularity $n$,
{\sf Count-Min} uses $w$ hash functions, whereas {\sf Equal-Sketch} employs $nw$ hash functions.
Since {\sf MOD-Sketch} optimally combines some modules of the key, the number of hash functions used
by it is less than $(n\times w)$, but more than $w$. {\em In all our experiments, the throughput of {\sf MOD-Sketch}
varies between 30K$\sim$90K items\ second}.
\vspace{-2mm}
\subsection{Generalizability}
\label{sec:general}
\vspace{-1mm}
We evaluate the generalizability of {\sf MOD-Sketch} by implementing it on top of the {\sf FCM} sketch \cite{TBAY09}.
{\sf FCM} improves the accuracy of {\sf Count-Min} by employing different number of hash functions for high-frequency
and low-frequency items, detected by an additional Misra-Gries counter \cite{MG82}. {\sf FCM} first applies two
separate hash functions to compute an \emph{offset} and a \emph{gap}, which determines the subset of hash functions
to be used for hashing the item. Then, it utilizes the selected hash functions
to hash the item into the sketch. We refer to our implementation of {\sf MOD-Sketch} on top of {\sf FCM}
as {\sf FMOD} (Figure~\ref{fig:FCMk16Top}). As expected, {\sf FCM} reduces the observed error
compared to {\sf Count-Min}. However, {\em our designed {\sf FMOD} further reduces the observed
error even compared to {\sf FCM}. These results demonstrate the generalizability of {\sf MOD-Sketch}}.
\vspace{-2mm}
\subsection{Effectiveness of Median Aggregate}
\begin{figure}[t!]
\vspace{-1mm}
\centering
\begin{minipage}{.45\linewidth}
  \includegraphics[scale=0.33]{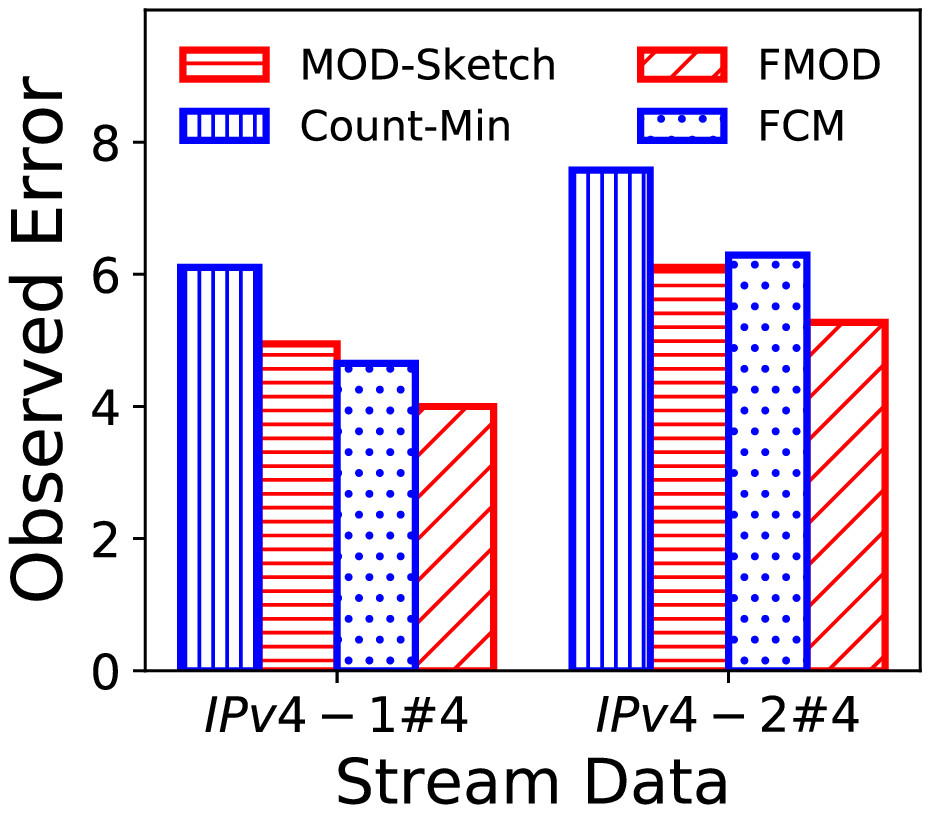}
  \vspace{-3mm}
  \captionof{figure}{\small Generality for {\sf MOD-Sketch}, coupled with {\sf FCM} sketch. Top-1000 query. Each sketch size is 320MB,
   with $h=4\times 10^{6}$, $w=10$.}
  \label{fig:FCMk16Top}
\end{minipage}
\hspace{.05\linewidth}
\begin{minipage}{.45\linewidth}
  \includegraphics[scale=0.33]{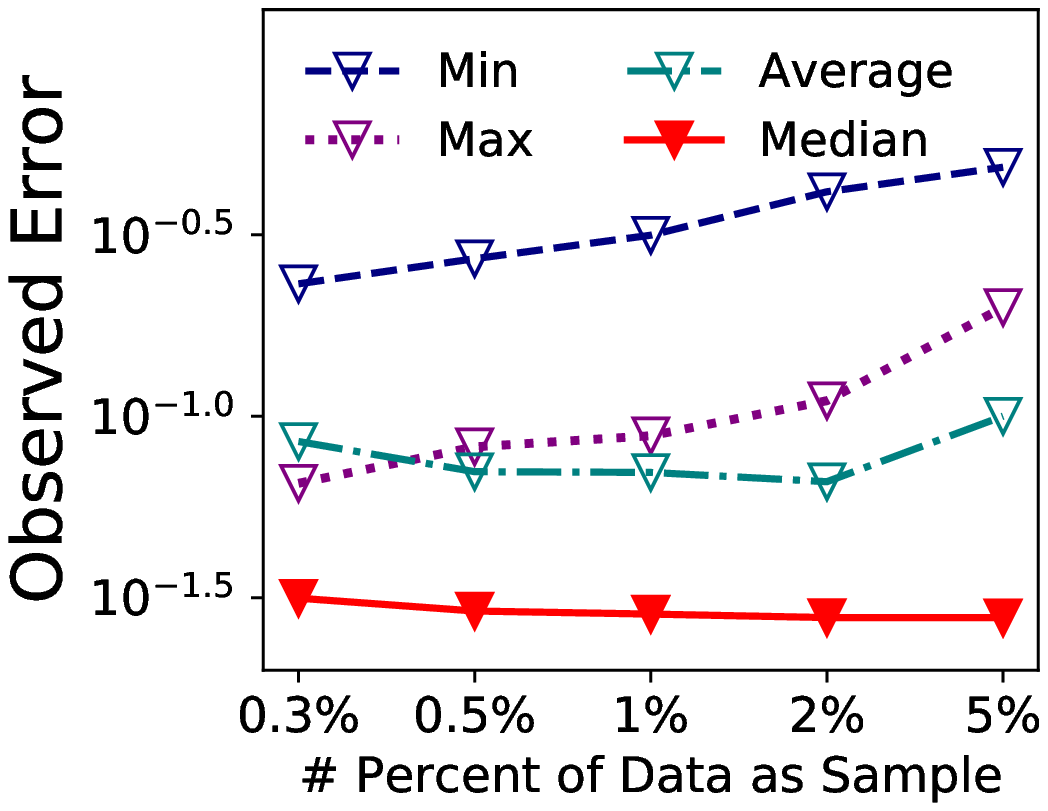}
  \vspace{-3mm}
  \captionof{figure}{\small Effectiveness of median aggregate in computing the optimal setting of {\sf MOD-Sketch}. Top-100 query, {\em Twitter}.}
  \label{fig:aggregateF}
\end{minipage}
\vspace{-6mm}
\end{figure}

\vspace{-1mm}
In Figure~\ref{fig:aggregateF}, we demonstrate the effectiveness of our median aggregate for selecting the parameter $\alpha$, compared to
other aggregates: minimum, maximum, and average. The median aggregate produces less observed error, this is because
the estimations from maximum and average (minimum) are greatly affected by a few sampled items with extremely large (small) frequency
in a skewed distribution. In contrast, the median estimate avoids such issues.

\vspace{-2mm}
\section{Conclusions}
\label{sec:conclusions}
\vspace{-1mm}
We present {\sf MOD-Sketch} to improve the accuracy of sketches
by employing multiple independent hash functions that hash different
modules in a key and their combinations separately. We develop
scalable algorithms that sample a small portion of the stream,
and find the optimal strategy to combine different modules of the
key before they are hashed into the sketch. Furthermore, we
compute good-quality hashing ranges for various combined
parts of the key. Based on our empirical results
over six real-world data streams, {\sf MOD-Sketch} outperforms several baselines
including {\sf Count-Min}, and it can be used together with
more sophisticated sketches, e.g., {\sf FCM} to further improve the frequency estimation
accuracy. In future work, it would be interesting to consider how to update {\sf MOD-Sketch}
parameters when the distribution of the items in the stream change over time.


{\scriptsize
\begin{thebibliography}{00}
\bibitem{Aggarwal03} C. Aggarwal, J. Han, J. Wang, and P. Yu, ``A Framework for Clustering Evolving Data Streams'', VLDB, 2003.
\bibitem{ADWR17} N. K. Ahmed, N. Duffield, T. L. Willke, and R. A. Rossi, ``On Sampling from Massive Graph Streams'', PVLDB, vol. 10, no. 11, pp. 1430--1441, 2017.
\bibitem{AMSz96} N. Alon, Y. Matias, and M. Szegedy, ``The Space Complexity of Approximating the Frequency Moments'', STOC, 1996.
\bibitem{Bloom} B. H. Bloom, ``Space/time Trade-offs in Hash Coding with Allowable Errors'', Comm. of ACM, vol. 13, issue 7, pp. 422--426,
1970.
\bibitem{C15} E. Cohen, ``Stream Sampling for Frequency Cap Statistics'', KDD, 2015.
\bibitem{CCF02} M. Charikar, K. Chen, and M. Farach-Colton, ``Finding Frequent Items in Data Streams'', ICALP, 2002.
\bibitem{CH08} G. Cormode and M. Hadjieleftheriou, ``Finding Frequent Items in Data Streams'', PVLDB,  vol. 1, no. 2, pp. 1530--1541, 2008.
\bibitem{CGMP12} G. Cormode and M. Garofalakis and P. J. Haas and C. Jermaine, ``Synopses for Massive Data: Samples, Histograms, Wavelets, Sketches'',
Foundations and Trends in Databases, vol. 4, no. 1--3, pp. 1--294, 2012.
\bibitem{countmin} G. Cormode and S. Muthukrishnan, ``An Improved Data-Stream Summary: The Count-min Sketch and its Applications'',
J. of Algorithms, vol. 55, no. 1, 2005.
\bibitem{CJKMSS04} G. Cormode, T. Johnson, F. Korn, S. Muthukrishnan, O. Spatscheck, and D. Srivastava, ``Holistic UDAFs at Streaming Speeds'',
SIGMOD, 2004.
\bibitem{DGGR02} A. Dobra, M. Garofalakis, J. Gehrke, and R. Rastogi, ``Processing Complex Aggregate Queries over Data Streams'', SIGMOD, 2002.
\bibitem{EV03} C. Estan and G. Varghese, ``New Directions in Traffic Measurement and Accounting: Focusing on the Elephants, Ignoring the Mice'',
ACM Trans. Comput. Syst., vol. 21, no. 3, pp. 270--313, 2003.
\bibitem{FSGMU98} M. Fang, N. Shivakumar, H. Garcia-Molina, R. Motwani, and J. Ullman, ``Computing Iceberg Queries Efficiently'', VLDB, 1998.
\bibitem{GG02} M. Garofalakis and P. B. Gibbons, ``Wavelet Synopses with Error Guarantees'', SIGMOD, 2002.
\bibitem{GGMS96} S. Ganguly, P. B. Gibbons, Y. Matias, and A. Silberschatz, ``Bifocal Sampling for Skew-Resistant Join Size Estimation'',
SIGMOD, 1996.
\bibitem{GKMS02} A. C. Gilbert, Y. Kotidis, S. Muthukrishnan, and M. Strauss, ``How to Summarize the Universe: Dynamic Maintenance of Quantiles'',
VLDB, 2002.
\bibitem{GKS01} S. Guha, N. Koudas, and K. Shim, ``Data-streams and Histograms'', STOC, 2001.
\bibitem{GSWS01} A. C. Gilbert, Y. Kotidis, S. Muthukrishnan, and M. Strauss, ``Surfing Wavelets on Streams: One-Pass Summaries for Approximate Aggregate Queries'', VLDB, 2001.
\bibitem{KA17} A. Khan and C. Aggarwal, ``Toward Query-Friendly Compression of Rapid Graph Streams'', Social Netw. Analys. Mining, vol. 7, no. 1,
pp. 23:1--23:19, 2017.
\bibitem{KSZC03} B. Krishnamurthy, S. Sen, Y. Zhang, and Y. Chen, ``Sketch-based Change Detection: Methods, Evaluation, and Applications'',
IMC, 2003.
\bibitem{MP09} N. Manerikar and T. Palpanas, ``Frequent Items in Streaming Data: An Experimental Evaluation of the State-of-the-art'',
Data Knowl. Eng., vol. 68, no. 4, pp. 415--430, 2009.
\bibitem{MAA05} A. Metwally, D. Agrawal, and A. E. Abbadi, ``Efficient Computation of Frequent and Top-k Elements in Data Streams'',
ICDT, 2005.
\bibitem{MG82} J. Misra and D. Gries, ``Finding Repeated Elements'', Science of Computer Programming, vol. 2, no. 2, pp. 143--152, 1982.
\bibitem{PTTW13} A. Pavan, K. Tangwongsan, S. Tirthapura, and K.-L. Wu, ``Counting and Sampling Triangles from a Graph Stream'', PVLDB, vol. 6, no. 14, pp. 1870--1881, 2013.
\bibitem{RKK14} O. Rottenstreich, Y. Kanizo, and I. Keslassy, ``The Variable-Increment Counting Bloom Filter'', IEEE/ACM Trans. Netw., vol. 22, no. 4, pp. 1092--1105, 2014.
\bibitem{RKA16} P. Roy, A. Khan, and G. Alonso, ``Augmented Sketch: Faster and More Accurate Stream Processing'', SIGMOD, 2016.
\bibitem{RD08} F. Rusu and A. Dobra, ``Sketches for Size of Join Estimation'', ACM TODS, vol. 33, no. 3, pp. 15, 2008.
\bibitem{SLCGGZDKM06} R. Schweller, Z. Li, Y. Chen, Y. Gao, A. Gupta, Y. Zhang, P. A. Dinda, M.-Y. Kao, and G. Memik, ``Reversible Sketches: Enabling Monitoring and Analysis over High-speed Data Streams'',
IEEE/ACM Trans. Netw., vol. 15, no. 5, pp. 1059--1072, 2007.
\bibitem{TCM16} N. Tang, Q. Chen, and P. Mitra, ``Graph Stream Summarization: From Big Bang to Big Crunch'', SIGMOD, 2016.
\bibitem{TBAY09} D. Thomas, R. Bordawekar, C. Aggarwal, and P. S. Yu, ``On Efficient Query Processing of Stream Counts on the Cell Processor'', ICDE, 2009.
\bibitem{WLKC16} J. Wang, W. Liu, S. Kumar, S.{-}F. Chang, ``Learning to Hash for Indexing Big Data - A Survey'', Proceedings of the IEEE, vol. 104, no. 1, pp. 34--57, 2016.
\bibitem{gsketch} P. Zhao, C. Aggarwal, and M. Wang, ``gSketch: On Query Estimation in Graph Streams'', PVLDB,  vol. 5, no. 3, pp. 193--204, 2011.
\bibitem{ZWS11} D. Zhang, F. Wang, L. Si, ``Composite Hashing with Multiple Information Sources'', SIGIR, 2011.
\bibitem{ZYJCYLU18} Y. Zhou, T. Yang, J. Jiang, B. Cui, M. Yu, X. Li, and S. Uhlig, ``Cold Filter: A Meta-Framework for Faster and More Accurate Stream Processing'', SIGMOD, 2018.
\end{thebibliography}
}

\end{document}